\documentclass[twocolumn]{aastex63}
\usepackage{natbib}
\usepackage{color} 
\usepackage{aas_macros} 
\usepackage{ulem}
\usepackage{hyperref} 
\usepackage{amsmath}
\revised{}
\submitjournal{AJ}

\shorttitle{density shapes of the GSE-related and the GSE-removed stellar halos}
\shortauthors{Wenbo et al.}
\begin{document}
{
\title{Influence of the Gaia-Sausage-Enceladus on the density shape of the Galactic stellar halo revealed by halo K giants from the LAMOST survey}

\correspondingauthor{Gang Zhao}
\email{gzhao@nao.cas.cn}
\author{Wenbo Wu}
\affil{CAS Key Laboratory of Optical Astronomy, National Astronomical Observatories, Chinese Academy of Sciences, Beijing 100101, People's Republic of China; gzhao@nao.cas.cn}
\affil{School of Astronomy and Space Science, University of Chinese Academy of Sciences, Beijing 100049, People's Republic of China}
\author{Gang Zhao}
\affil{CAS Key Laboratory of Optical Astronomy, National Astronomical Observatories, Chinese Academy of Sciences, Beijing 100101, People's Republic of China; gzhao@nao.cas.cn}
\affil{School of Astronomy and Space Science, University of Chinese Academy of Sciences, Beijing 100049, People's Republic of China}
\author{Xiang-Xiang Xue} \affil{CAS Key Laboratory of Optical Astronomy, National Astronomical
  Observatories, Chinese Academy of Sciences, Beijing 100101, People's Republic of China; gzhao@nao.cas.cn} \affil{School of Astronomy and Space Science, University of Chinese Academy of Sciences, Beijing 100049, People's Republic of China}
\author{Wenxiang Pei}
\affil{Key Laboratory for Computational Astrophysics, National Astronomical Observatories, Chinese Academy of Sciences, Beijing 100101, China}
\affil{School of Astronomy and Space Science, University of Chinese Academy of Sciences, Beijing 100049, People's Republic of China}
\author{Chengqun Yang}
\affiliation{Shanghai Astronomical Observatory, Chinese Academy of Sciences, 80 Nandan Road, Shanghai 200030, People’s Republic of China}



\begin{abstract} \label{abstract}
We present a study of the influence of the Gaia-Sausage-Enceladus (GSE) on the density shape of the Galactic stellar halo using 11624 K giants from the LAMOST survey. Every star is assigned a probability of being a member of the GSE based on its spherical velocities and metallicity by a Gaussian Mixture Model. We divide the stellar halo into two parts by the obtained probabilities, of which one is composed of the GSE members and defined as the GSE-related halo, and the other one is referred to as the GSE-removed halo. Using a non-parametric method, the radial number density profiles of the two stellar halos can be well described by a single power law with a variable flattening $q$ ($r = \sqrt{R^2+[(Z/q(r))]^2}, \nu = {\nu_0}r^{-\alpha}$). The index $\alpha$ is $4.92\pm0.12$ for the GSE-related halo and $4.25\pm0.14$ for the GSE-removed halo. Both the two stellar halos are vertically flattened at smaller radii but become more spherical at larger radii. We find that the GSE-related halo is less vertically flattened than the GSE-removed halo, and the difference of $q$ between the two stellar halos ranges from 0.07 to 0.15. However, after the consideration of the bias, it is thought to be within 0.08 at most of the radii. Finally, we compare our results with two Milky Way analogues which experience a significant major merger in the TNG50 simulation. The study of the two analogues also shows that the major merger-related stellar halo has a smaller ellipticity than the major merger-removed stellar halo.  
\end{abstract}


\section{Introduction} \label{sec:intro}

In the $\Lambda$ cold dark matter ($\Lambda$ CDM) cosmological model, the galaxy structure forms hierarchically: galaxies grow by accretion from the mergers of smaller satellite galaxies \citep{1978MNRAS.183..341W,peebles2020large}. According to the stellar mass ratio $\mu$ of the satellite to the host galaxy, mergers can be categorized into major ($\mu \geq \frac{1}{4}$), minor($\frac{1}{4} \ge \mu \geq \frac{1}{10}$), and very minor ($\frac{1}{10} \geq \mu$) types. These mergers play an important role in driving the growth of the stellar components and determining their chemodynamical properties. 
With the advent of the large sky surveys, our understanding of the formation and evolution of the Galactic stellar halo has steadily advanced over the past decades. Recent studies showed that the bulk of the accreted stellar halo is contributed by several massive dwarf galaxies. Observational evidence of the merger events can be found in the stellar kinematics \citep{2004A&A...418..989N,2009ApJ...692L.113Z,2014ApJ...787...31Z,2017ApJ...844..152L,2018ApJ...862L...1D,2019NatAs...3..932G,2020MNRAS.497..109F,2020ApJ...903...25N,2021SCPMA..6439562Z,2022ApJ...924...23W}, chemical abundance patterns \citep{2020MNRAS.493.5195D,2021MNRAS.508.1489F,2021MNRAS.tmp.3190B,2021ApJ...908L...8A}, halo substructures \citep{2016ApJ...816...80J,2019ApJ...880...65Y,2019A&A...631L...9K,2020ApJ...891...39Y,2020ApJ...901...48N,2021ApJ...908..191C}, and globular clusters brought by the satellites \citep{2018ApJ...863L..28M,2019MNRAS.486.3180K,2019MNRAS.488.1235M,2021AJ....162...42W}. The merger process with a massive dwarf galaxy not only leaves observational relics like a local tidal stream but also could affect the overall morphology of the stellar halo \citep{2013ApJ...763..113D,2015ApJ...809..144X,2016MNRAS.458.2371R,2019MNRAS.485.2589M}. Therefore, understanding the spatial distribution of halo stars, specifically the radial density profile and vertical flattening, is critical to disentangle the mass assembly history of the Milky Way.


 Early studies attempting to measure the global structure of the Milky Way stellar halo had to rely on a small number of tracers and/or sky coverage, and they usually described the radial density profile by a single power law (SPL) with an index of $2.5-3.5$ and a constant flattening $q$ of $0.5-1$ \citep{1976AJ.....81.1095H,1984MNRAS.206..433H,1987MNRAS.227P..21S,1991ApJ...375..121P,1996AJ....112.1046W,1998MNRAS.298..387D}. Recently, with the advent of large sky surveys, the density shape of the stellar halo can be measured out to a large distance using a variety of stellar tracers with an expanding number and sky coverage. Several studies seemed to favor a broken power law (BPL) with constant flattening using different tracers of main-sequence turn off stars \citep{2011ApJ...731....4S,2015A&A...579A..38P}, blue straggler and horizontal branch stars \citep{2011MNRAS.416.2903D,2016MNRAS.463.3169D}, RR Lyraes \citep{2009MNRAS.398.1757W,2010ApJ...708..717S,2014ApJ...788..105F,2021MNRAS.tmp.3392L}. They found a break radius $r_\mathrm{break}$ in between 19 and 30 kpc. Stars inside $r_\mathrm{break}$ followed a power law with an index of about $2.5$, while stars outside $r_\mathrm{break}$ followed a much steeper one with an index of approximately $4$. However, the existence of $r_\mathrm{break}$ is controversial. \citet{xue2015radial} found that the radial density profile of the SEGUE K giants can be fitted well by a BPL with constant flattening, or a SPL with variable flattening alternatively. Further studies also showed that a break radius is not necessarily needed assuming that the halo density shape varies with radius, and the index $\alpha$ is about $4-5$ using an oblate ellipsoid model \citep{2016MNRAS.463.3169D,2018ApJ...859...31H,2018MNRAS.473.1244X}, or $2.96$ using a triaxial ellipsoid model \citep{2018MNRAS.474.2142I}. 

In this study, we aim to explore the influence of the Gaia-Sausage-Enceladus on the density shape of the stellar halo using 11624 halo K giants selected from the LAMOST survey. The data release of the \textit{Gaia} satellite has contributed significantly to our understanding of the Galactic formation history. One of the most insightful findings is that our galaxy had experienced a major merger event with a dwarf galaxy named as the Gaia-Sausage-Enceladus (GSE) \citep{2018MNRAS.478..611B,2018Natur.563...85H}. The stellar debris of the GSE is characterized by the large eccentricity (ec) and low angular momentum along the z-axis ($L_\mathrm{z}$), and the metallicity distribution function of the GSE members has a peak of [Fe/H] ranging from $-1.4$ to $-1.2$ dex \citep{2019ApJ...881L..10S,2019NatAs...3..932G,2020MNRAS.493.5195D,2020MNRAS.497..109F,2020ApJ...901...48N,2020MNRAS.492.3631M}. By studying the chemodynamic properties of its possible members, the GSE is thought to be a large dwarf galaxy with a stellar mass of the order of $10^9 - 10^{10}\,M_\odot$\citep{2019MNRAS.487L..47V,2019MNRAS.490.3426D,2019MNRAS.484.4471F,2019MNRAS.482.3426M}. The head-on collision of the ancient GSE and the Milky Way is expected to happen around $8-10$ Gyr ago \citep{2019ApJ...881L..10S,2019ApJ...883L...5B,2020ApJ...897L..18B}. 

The stellar debris of the GSE is first identified in the solar neighborhood, and then shown to stretch from the inner to the outer stellar halo \citep{2019MNRAS.486..378L,2021ApJ...919...66B,2022ApJ...924...23W}. Using different tracers of blue horizontal branch (BHB) stars, K giants, main sequence stars, and RR Lyraes, recent studies found that the GSE could deliver over half of the Galactic stellar halo \citep{2019MNRAS.486..378L,2019ApJ...874....3N,2021MNRAS.502.5686I,2021ApJ...918...74A,2022ApJ...924...23W,2022ApJS..258...20A}. Considering the large fraction of stars originating from the GSE, the global structure of the Galactic stellar halo is expected to be heavily influenced by this merger event. The influence of the major merger on the morphology of the host galaxy including the dark matter and stellar halos has been widely studied in many previous simulation works \citep{2003MNRAS.339...12Z,2011MNRAS.416.1377V,2018MNRAS.479.4004E,2019MNRAS.487..993D,2019MNRAS.485.2589M,2021A&A...647A..95P}. The galaxy halo is elongated along the merger axis, and become less concentrated due to the last major merger. \citet{2019MNRAS.487..993D} found that mergers on radial orbits lead to a prolate dark matter halo, while mergers on tangential orbits produce oblate remnants. \citet{2021A&A...647A..95P} studied the stellar halo morphology of 1114 early type galaxies in TNG50 simulation. They also found that galaxies with a larger fraction of stars originating from the major merger events tend to be less vertically flattened on average. Although it is hard to trace the evolution of the Galactic halo from the beginning of the major merger, the present density shapes of the accreted and in-situ stellar halos may indicate the influence of the GSE. By studying the orbits of local metal poor stars, \citet{2021arXiv210810525S} showed that the global structure of the subsample of stars kinematically associated with the GSE is more spherical with $q \sim 0.9$ than the general halo sample. Therefore, the Galactic stellar halo may evolve to be less vertically flattened due to the last major merger of the GSE. 

This paper is organized as follows. In Section~\ref{sec:data}, we introduce the observational data of halo K giants used in this study. Then we describe the Gaussian Mixture model (GMM) and apply it to the observational data to select the possible GSE members in Section~\ref{sec:method}. The non-parametric method used for estimating the radial density profile is presented in Section~\ref{sec:fit}. We analyze the density shapes of the GSE-related and the GSE-removed halos in Section~\ref{sub:real}, and compare our results with two Milky Way analogues in TNG50 simulation in Section~\ref{sub:simulation}. Finally, conclusions are made in Section~\ref{sec:summary}.

\section{Observational Data} \label{sec:data}
Halo K giants used in this study are selected from the LAMOST (Large Sky Area Multi-Object Fiber Spectroscopic
Telescope) DR5 catalog \citep{2006ChJAA...6..265Z,2012RAA....12..723Z,2012RAA....12.1197C,2012RAA....12.1243L,2015RAA....15.1089L}. These K giants are identified by a support vector machine classifier based on the spectral line features \citep{2014ApJ...790..110L}. As shown in \citet{2014ApJ...790..110L}, the contamination by stars other than K giants is smaller than 2.5\%. Distances of those K giants are estimated based on the fiducial isochrones using a Bayesian method presented in \citet{2014ApJ...784..170X}. The photometric information used for correcting the selection effects is publicly released in the 2Mass (Two Micron All-Sky Survey) \citep{2006AJ....131.1163S}. Line-of-sight velocity and stellar metallicity are derived by the well-calibrated LAMOST 1D pipeline \citep{2011RAA....11..924W,2014IAUS..306..340W}. Proper motions are taken from \textit{Gaia} EDR3 \citep{2021A&A...649A...1G} by cross-matching with a radius of $1^{\prime \prime}$. The stellar sample provides 7D information for the halo stars, including equatorial coordinate information ($\alpha$, $\delta$), heliocentric distance d, line-of-sight velocities $v_\text{los}$, proper motions ($\mu_{\alpha}$, $\mu_{\delta}$), and stellar metallicity [Fe/H]. The selected halo K giants have a mean distance precision ($\delta d/d$) of 13\%, mean $v_\text{los}$ uncertainty of 7 km $\textup{s}^{-1}$, and mean metallicity error of 0.14 dex. The proper motion uncertainties of 95\% of these halo stars are within 0.10 mas $\textup{yr}^{-1}$. 

The 7D information is transformed to the Galactocentric Cartesian coordinates ($X, Y, Z$) and velocities ($U, V, W$) using the conventions in \texttt{astropy} \citep{2018AJ....156..123A}. In Galactocentric Cartesian coordinates, we use the default values of the Solar Galactocentric distance $r_{\mathrm{gc},\odot}$ = 8.122 kpc \citep{2018A&A...615L..15G}, and height $Z_{\odot}$ = 20.8 pc \citep{2019MNRAS.482.1417B}. We adopt a Solar motion of (+12.9, +245.6, +7.78) km $\textup{s}^{-1}$ \citep{2004ApJ...616..872R,2018A&A...615L..15G,2018RNAAS...2..210D}. To transform to the Galactocentric spherical coordinates (\textit{$r$, $\theta, \phi$}) and velocities ($v_\mathrm{r}, v_{\theta}, v_{\phi}$), we follow the Equations 1$-$6 in \citet{2022ApJ...924...23W}. The uncertainties in distance, line-of-sight radial velocity, and proper motion (including the measurement error and covariance of proper motions) are propagated using a Monte Carlo sampling to estimate the median, standard error, and covariance of the velocities ($v_{r}, v_{\theta}, v_{\phi}$). 

We first select stars with a relative distance precision better than 30\%, and retain stars with $|Z| > 2$ kpc and [Fe/H] $< -0.5$ to exclude possible disk stars. To eliminate egregious outliers, we require the absolute spherical velocities to be $<$ 400 km $\text{s}^{-1}$ and the uncertainties $<$ 150 km $\text{s}^{-1}$. 

\citet{2022ApJ...924...23W} defined the stellar halo with the Sagittarius (Sgr) stream members removed as the Sgr-removed stellar halo. The unrelaxed Sgr stream \citep{1994Natur.370..194I} is the most obvious substructure in the Galactic halo, which may cause some bias in the GMM fitting \citep{2019MNRAS.486..378L,2022ApJ...924...23W}. To obtain an unbiased measurement of the velocity ellipsoid and metallicity distribution, we remove the Sgr stream members published by \citet{2019ApJ...886..154Y}, which contains about 1700 K giants selected from the LAMOST DR5. A detailed description of the integrals of motion method used for selecting the Sgr stream members can be found in X.X-Xue et al. (2021, in preparation) and \citet{2022ApJ...924...23W}. We use a less strict cut of $|Z|$ than \citet{2022ApJ...924...23W}, in order to reduce the possible bias caused by the cut of $|Z|$ in estimating the stellar density. Figure~\ref{fig:spatial} shows the spatial distributions (on $\sqrt{X^2+Y^2}$-$Z$, $X-Y$ and $\textit{l}-\textit{b}$ planes) of our K giants, where most of them are located within $r_\mathrm{gc}$ of 35 kpc. In the following words, we will divide these halo K giants into two components by the GMM, of which one is composed of the possible GSE members and referred to as the GSE-related halo, and the other one is defined as the GSE-removed halo. The exploration of the difference in density shapes between the GSE-related and the GSE-removed stellar halos will help us further understand the influence of the GSE on the global structure of the Galactic stellar halo.

\begin{figure*}
	\centering
	\includegraphics[width=1.0\textwidth]{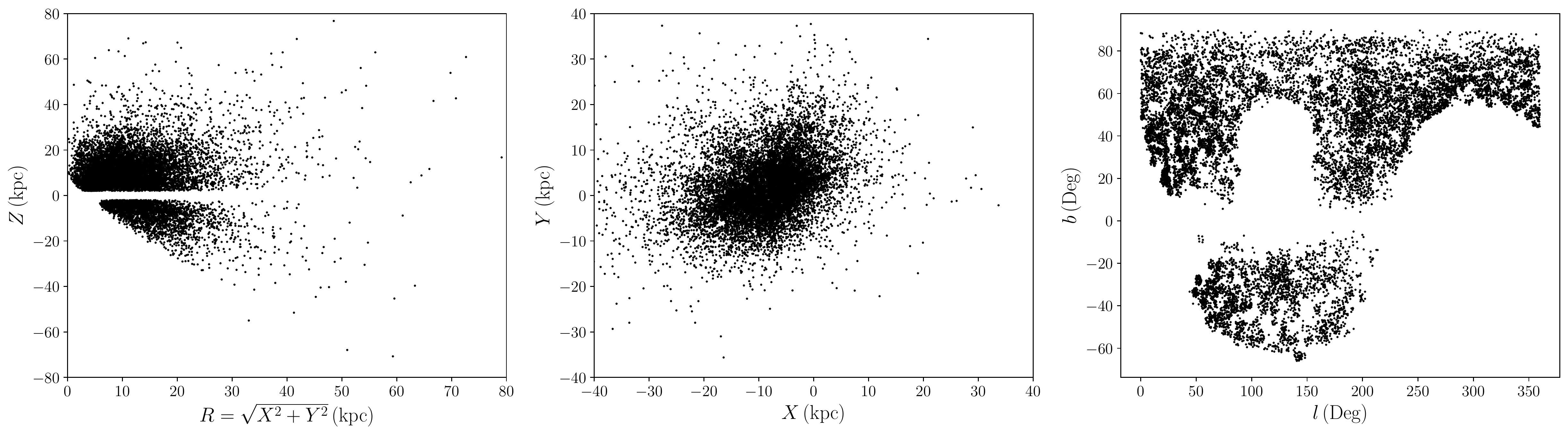}
	\caption{Spatial distributions of $R-Z$ (left), $X-Y$ (middle), and $\textit{l}-\textit{b}$ (right) for the LAMOST halo K giant sample.}
	\label{fig:spatial}
\end{figure*}

\section{Selection Methods and Results} \label{sec:method}
In this section, we divide the Galactic stellar halo into two parts of the GSE-related and the GSE-removed halos by the GMM. The distributions of $L_\mathrm{z}$ and ec are shown for the two selected halos. We apply the GMM to the mock data to check the accuracy of the selection in Section~\ref{sub:test}.

\subsection{Gaussian Mixture model}\label{sub:gmm}

The Galactic stellar halo is thought to be heavily influenced by the ancient major merger of the GSE. To reveal the contribution of the GSE to the stellar halo, recent studies applied a Gaussian Mixture Model to describe the distributions of the spherical velocities and metallicity of halo stars \citep{2019MNRAS.486..378L,2019ApJ...874....3N,2021MNRAS.502.5686I,2021ApJ...918...74A,2022ApJ...924...23W}. The GMM divides the stellar halo into two components, of which one is a more metal poor and isotropic stellar halo referred to as the GMM/isotropic component, and the other one represents a more metal rich and radially biased stellar halo named as the GMM/anisotropic or GMM/Sausage component. In \citet{2022ApJ...924...23W}, we explore the contribution of the GSE to the stellar halo by making use of the GMM and applying it to halo star samples of LAMOST K giants, SEGUE K giants \citep{2014ApJ...784..170X}, and SDSS BHB stars \citep{2011ApJ...738...79X}. We will briefly introduce the GMM hereafter.

For a star $D_i (v_{r,i}, v_{\theta,i}, v_{\phi,i}, \mathrm{[Fe/H]}_i)$, the likelihoods of the isotropic and anisotropic stellar halos are shown in Equation~\ref{eq:iso} and Equation~\ref{eq:ani},

\begin{equation}
\mathcal{L}_\mathrm{iso}(D_i|\theta) = \mathcal{N}(\boldsymbol{v_i}|\mathbf{0}, \Sigma_{i}^\mathrm{iso})\mathcal{N}(\text{[Fe/H]}_{i}|\mu_\mathrm{[Fe/H]}^\mathrm{iso}, {\sigma_{\mathrm{[Fe/H]},i}^\mathrm{iso}}^2)
\label{eq:iso}
\end{equation}

\begin{equation}
\begin{aligned}
\mathcal{L}_\mathrm{an}(D_i|\theta) = \frac{1}{2}[\mathcal{N}(\boldsymbol{v_i}|\boldsymbol{V}^{\widetilde{\mathrm{an}}}, \Sigma_{i}^\mathrm{an}) + \mathcal{N}(\boldsymbol{v_i}|\boldsymbol{V}^\text{an}, \Sigma_{i}^\text{an})]\\
\times\mathcal{N}(\text{[Fe/H]}_{i}|\mu_\mathrm{[Fe/H]}^\mathrm{an}, {\sigma_{\text{[Fe/H]},i}^\mathrm{an}}^2),
\end{aligned}
\label{eq:ani}
\end{equation}
where $\boldsymbol{v_i}$ and $\text{[Fe/H]}_{i}$ are the Galactocentric spherical velocity $(v_{r,i}, v_{\theta,i}, v_{\phi,i})$ and the stellar metallicity of the star $i$ respectively.

In Equation~\ref{eq:iso}, the spherical velocities of the isotropic halo are described by a three dimensional single Gaussian with mean spherical velocities set to be zero. We enforce the velocity tensor in the tangential direction to be isotropic by setting $\sigma_{v_\theta;\mathrm{iso}} = \sigma_{v_\phi;\mathrm{iso}}$, and $\sigma_{t;\mathrm{iso}}$ is used as a unified representation of the tangential velocity dispersion. The velocity dispersion $\Sigma_{i}^\mathrm{iso}$ is a combination of the covariance matrix of measurement error $\Sigma_i$ = diag$(\sigma_{v_r;i}^2, \sigma_{v_\theta;i}^2, \sigma_{v_\phi;i}^2)$ and the intrinsic dispersion $\Sigma_\mathrm{iso}$ =  diag$(\sigma_{v_r;\mathrm{iso}}^2, \sigma_{t;\mathrm{iso}}^2, \sigma_{t;\mathrm{iso}}^2)$. We define $\mu_\mathrm{[Fe/H]}^\mathrm{iso}$ as the mean metallicity. The metallicity dispersion $\sigma_{\text{[Fe/H]},i}^\text{iso}$ is a combination (in quadrature) of the individual measurement error $\sigma_{\text{[Fe/H]},i}$ and the intrinsic dispersion $\sigma_{\text{[Fe/H]}}^\text{iso}$ ($\sigma_{\text{[Fe/H]};i}^\mathrm{iso} = \sqrt{{\sigma_{\text{[Fe/H]},i}}^2 + {\sigma_{\text{[Fe/H]}}^\mathrm{iso}}^2}$).


In Equation~\ref{eq:ani}, the spherical velocities of the anisotropic halo are described by a symmetric bimodal Gaussian with two lobes of high positive and negative radial velocities. To track the dynamical state of the stellar halo, \citet{2018MNRAS.478..611B} fitted the shape of the velocity ellipsoid of SDSS-\textit{Gaia} main sequence stars with a zero-mean multi-variate Gaussian model. They found that the residuals show clear over-densities of stars with high positive and negative radial velocity, especially in the most metal rich stellar halo. The two distinct high $v_r$ lobes are proven to be the stellar debris of the GSE. We assume that the GMM/Sausage component is built from two Gaussians with an equal mixing fraction. To match the two $v_r$ lobes, the mean radial velocities of the two Gaussians are defined as $+\langle v_{r}^\mathrm{an} \rangle$ and $-\langle v_{r}^\mathrm{an} \rangle$. The mean rotation velocity $\langle v_\phi^\mathrm{an} \rangle$ of the \textit{Gaia}-Sausage is not forced to be zero based on the analyses of \citet{2018MNRAS.478..611B}, \citet{2018Natur.563...85H}, and \citet{2019MNRAS.486..378L}. In the two Gaussians, the mean velocity $\boldsymbol{V}^{\text{an}}$ is set to be $(\langle v_{r}^\mathrm{an} \rangle, 0, \langle v_\phi^\mathrm{an} \rangle)$, and $\boldsymbol{V}^{\widetilde{\text{an}}}$ is $(-\langle v_r^\mathrm{an} \rangle, 0, \langle v_\phi^\mathrm{an} \rangle)$. The velocity dispersion of the GMM/Sausage component $\Sigma_{i}^\text{an}$ is a combination of the covariance matrix of measurement error $\Sigma_{i}$ and the intrinsic dispersion $\Sigma_\mathrm{an}$ =    
diag$(\sigma_{v_r;\text{an}}^2, \sigma_{t;\text{an}}^2, \sigma_{t;\text{an}}^2)$. We define $\mu_\mathrm{[Fe/H]}^\mathrm{an}$ as the mean metallicity of the anisotropic stellar halo. The full metallicity dispersion $\sigma_{\text{[Fe/H]},i}^\text{an}$ is a combination (in quadrature) of the measurement error $\sigma_{\text{[Fe/H]},i}$ and the intrinsic dispersion $\sigma_{\text{[Fe/H]}}^\text{an}$.

The likelihood of observing one star $D_i (v_{r,i}, v_{\theta,i}, v_{\phi,i}, \mathrm{[Fe/H]}_i)$ is defined as,
\begin{equation}
\begin{aligned}
\mathcal{L}(D_i|\theta) = f_\mathrm{iso}\mathcal{L}_\mathrm{iso}(D_i|\theta) + f_\mathrm{an}\mathcal{L}_\mathrm{an}(D_i|\theta),
\end{aligned}
\label{eq:GMM}
\end{equation}
where $f_\mathrm{iso}$ is the contribution of the isotropic component, and $f_\mathrm{an}$ refers to the fraction of the anisotropic component, such that $f_\mathrm{iso} + f_\mathrm{an} = 1$.

There are 11 free parameters in the GMM. The likelihood of the GMM for a halo sample containing $N$ stars is defined as,
\begin{equation}
\begin{aligned}
\mathcal{L}(D|\theta) = \prod_{i=1}^{N}\mathcal{L}(D_i|\theta).
\end{aligned}
\label{eq:GMM_all}
\end{equation}

We allocate the K giants to different $r_\mathrm{gc}$ bins and apply the GMM to each bin to obtain the best estimated parameters. To avoid the possible over-fitting caused by the small number of stars in the outer halo ($r_\mathrm{gc} > 30$ kpc), we concatenate the two $r_\mathrm{gc}$ bins in \citet{2022ApJ...924...23W} together and use only one $r_\mathrm{gc}$ bin in Table~\ref{tab:Sgr-removed}. The MCMC fitting is accomplished by \texttt{Emcee}, which is an implementation of Goodman \& Weare’s Affine Invariant Markov chain Monte Carlo Ensemble sampler \citep{2013PASP..125..306F}. We use 200 walkers and 1000 steps as burn-in, then followed by 3000 steps. We define the 50th percentile of the marginalized posterior distributions as the best estimated value and the 16th and 84th as the uncertainties. The best estimated parameters shown in Table~\ref{tab:Sgr-removed} are almost the same as the results of \citet{2022ApJ...924...23W}. 

In a $r_\mathrm{gc}$ bin, we define the probability of a star $D_i$ belonging to the GSE as $\mathrm{prob(an)}_i$ in Equation~\ref{eq:best},
\begin{equation}
\mathrm{prob(an)}_i = \frac{f_\mathrm{an}\mathcal{L}_\mathrm{an}(D_i|\theta)}{f_\mathrm{iso}\mathcal{L}_\mathrm{iso}(D_i|\theta) + f_\mathrm{an}\mathcal{L}_\mathrm{an}(D_i|\theta)},
\label{eq:best}
\end{equation}
where the model parameters ($\theta, f_\mathrm{an}, f_\mathrm{iso}$) are the best estimated parameters of the corresponding $r_\mathrm{gc}$ bin. We select stars with relatively larger $\mathrm{prob(an)}$ from this bin according to the fraction of $f_\mathrm{an}$. We collect the selected stars in all $r_\mathrm{gc}$ bins together and define them as the GSE-related halo, and the other stars are defined as the GSE-removed halo. As shown in Figure~\ref{fig:compare_related_removed}, the main difference between the two stellar halos is reflected in the distributions of the tangential velocities and metallicity, which is consistent with the chemodynamic properties of the GSE as a highly radially biased and relatively more metal rich subcomponent. Figure~\ref{fig:prob} shows a clear division in the distributions of prob(an) between the two stellar halos. Most of the GSE-related halo stars have a prob(an) larger than 0.8, while prob(an) of 70\% of the GSE-removed halo stars is smaller than 0.4. We apply the GMM to the GSE-related halo again to check the possible residual from the GMM/isotropic halo. The large $f_\mathrm{an}$ ($> 0.99$) suggests a negligible contamination of the stars from the GMM/isotropic halo. 

Previous studies usually select the possible GSE members from halo stars by a robust kinematic cut of  $|{L_z}| < 500\,\mathrm{kpc\,km\,s^{-1}}$ or $\mathrm{ec} > 0.7$. Although the majority of the GSE stars have a large eccentricity and small $|{L_z}|$, there is still a certain amount of them located outside of the selection criteria as revealed by the \textit{N}-body simulation of the GSE stellar debris \citep{2021ApJ...923...92N}. We show the normalized distributions of $|{L_z}|$ of the two stellar halos in Figure~\ref{fig:lz}. The stellar debris of the simulated GSE is added as a comparison. The fractions of stars satisfying $|{L_z}| > 500$ are 0.73 for the GSE-related halo, 0.77 for the simulated GSE, and 0.28 for the GSE-removed halo. Our selected GSE-related halo is consistent with the simulated GSE in the distribution of $L\mathrm{z}$. Figure~\ref{fig:ec} shows the number distributions of the eccentricity of the two stellar halos. We find a division at ec = 0.67, beyond which the GSE-related halo stars become the dominated part of the stellar halo. 

\begin{figure*}
	\includegraphics[width=1\textwidth]{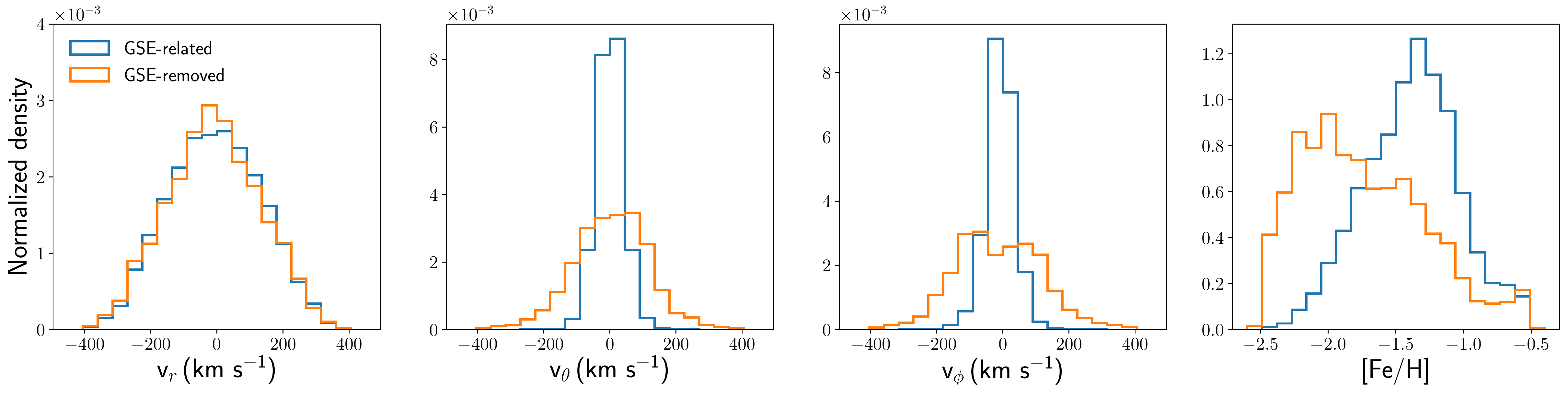}
	\caption{Histograms of the normalized density distributions of the spherical velocities and metallicity of the GSE-related (blue) and GSE-removed (orange) halos.}
	\label{fig:compare_related_removed}
\end{figure*}

\begin{figure}
	\centering
	\includegraphics[width=0.5\textwidth]{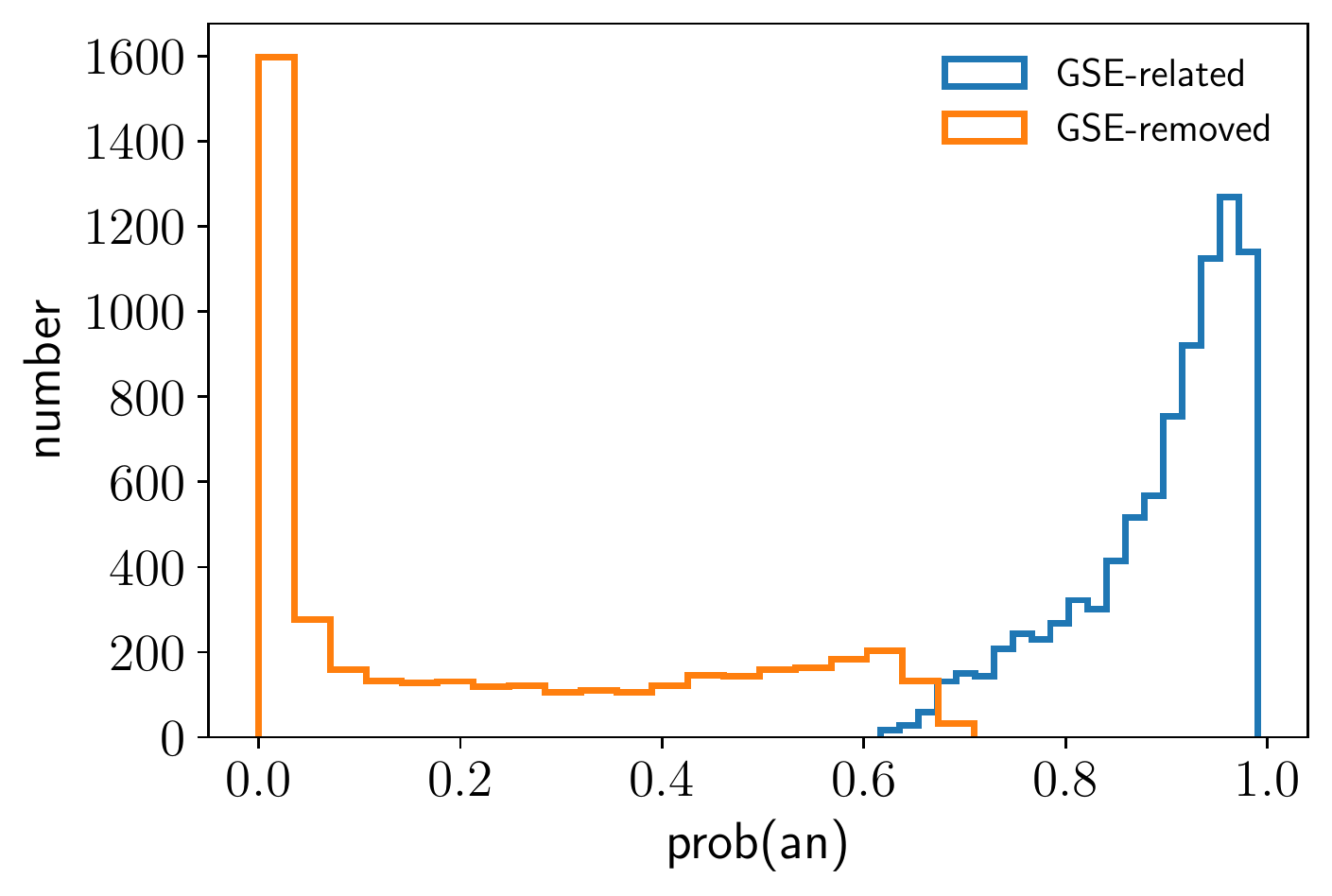}
	\caption{Histograms of the number distributions of prob(an) of the GSE-related (blue histogram) and GSE-removed (orange histogram) halos. Most of K giants of the GSE-related halo have a prob(an) larger than 0.85, while prob(an) of 70\% of K giants of the GSE-removed halo is smaller than 0.4. }
	\label{fig:prob}
\end{figure}

\begin{figure}
	\centering
	\includegraphics[width=0.5\textwidth]{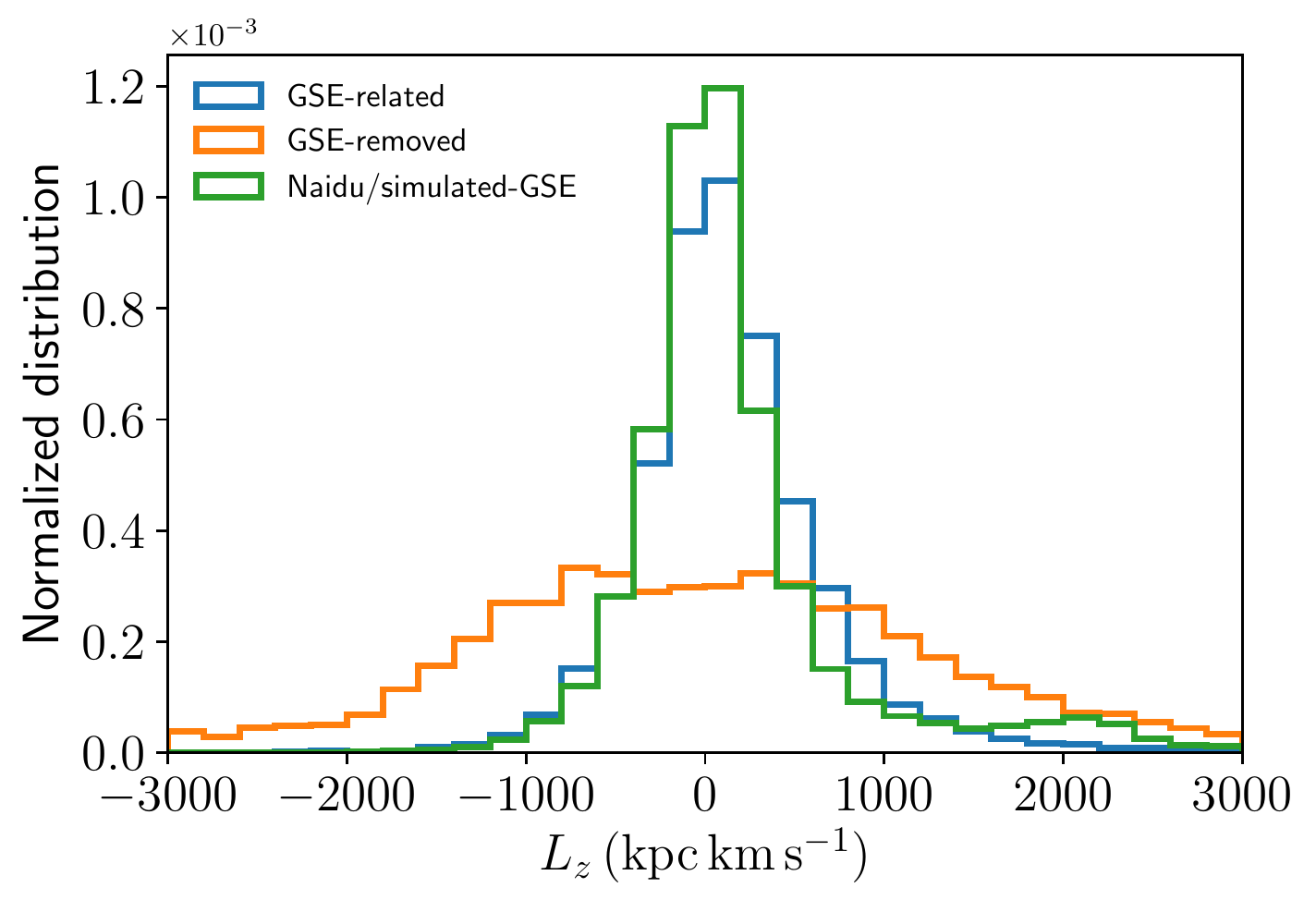}
	\caption{Histograms of the normalized distributions of the angular momentum ${L_z}$ for stars of the GSE-related halo (blue), GSE-removed halo (orange), and the stellar debris of the simulated GSE of \citet{2021ApJ...923...92N} (green). The GSE-related halo has a much stronger distribution in $|{L_z}| < 500\,\mathrm{kpc\,km\,s^{-1}}$ than the GSE-removed halo, which conforms to the highly radial orbits of the stellar debris of the GSE. However, there is still a certain amount of stars with $|{L_z}| > 500\,\mathrm{kpc\,km\,s^{-1}}$ both in the observational and simulation data of the GSE. A robust cut of ${L_z}$ may lead to a deficiency of them.}  
	\label{fig:lz}
\end{figure}


\begin{figure}
	\centering
	\includegraphics[width=0.5\textwidth]{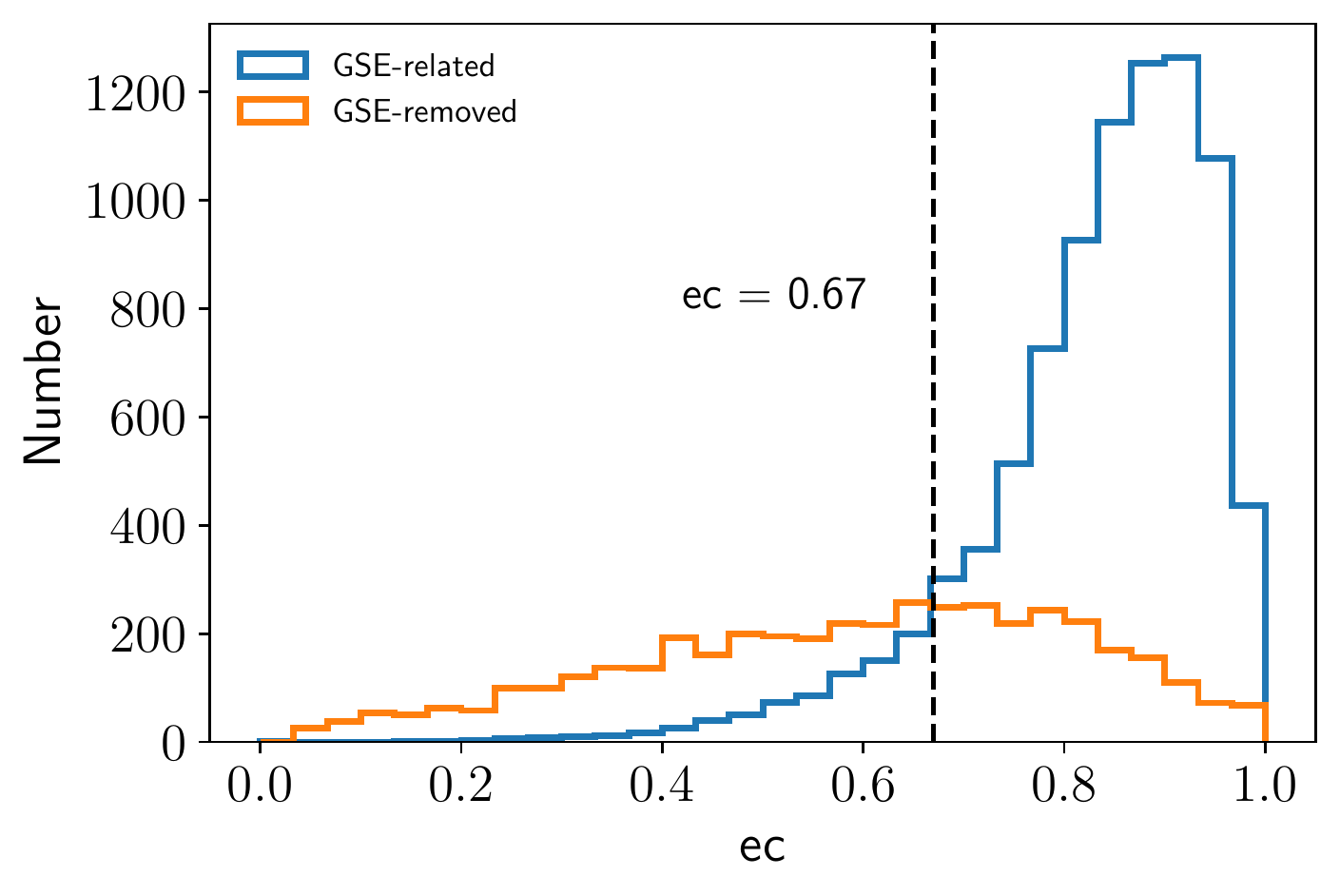}
	\caption{Histograms of the number distributions of eccentricity ec for stars of the GSE-related (blue) and GSE-removed (orange) halos. We find a division in eccentricity at ec = 0.67, where number of stars in either halos is equal.}
	\label{fig:ec}
\end{figure}

\begin{deluxetable*}{ccccccccccccc}
	\label{tab:Sgr-removed}
	\centering
	\tablecaption{Best estimated parameters of the GMM for the Sgr-removed LAMOST halo K gaint sample.}
	\tablewidth{0pt}
	\tablehead{
		\colhead{$r_\mathrm{gc}^\mathrm{a}$}&\colhead{$N^\mathrm{b}$}&\colhead{$\langle v_{r}^\mathrm{an} \rangle$}&\colhead{$\langle v_\phi^\mathrm{an} \rangle$}&\colhead{$\sigma_{v_r;\mathrm{an}}$}&\colhead{$\sigma_{t;\mathrm{an}}$}&\colhead{$\sigma_{v_r;\mathrm{iso}}$}&\colhead{$\sigma_{t;\mathrm{iso}}$}&\colhead{$\mu_\text{[Fe/H]}^\mathrm{an}$}&\colhead{$\sigma_\text{[Fe/H]}^\mathrm{an}$}&\colhead{$\mu_\text{[Fe/H]}^\mathrm{iso}$}&\colhead{$\sigma_\text{[Fe/H]}^\mathrm{iso}$}&\colhead{$f_\mathrm{an}$}\\
		\colhead{kpc}&\colhead{}&\colhead{km $\text{s}^{-1}$}&\colhead{km $\text{s}^{-1}$}&\colhead{km $\text{s}^{-1}$}&\colhead{km $\text{s}^{-1}$}&\colhead{km $\text{s}^{-1}$}&\colhead{km $\text{s}^{-1}$}&\colhead{dex}&\colhead{dex}&\colhead{dex}&\colhead{dex}}
	\decimals
	\startdata
	\hline
	$~\,2.00-~\,9.60$&2281&$114^{+6}_{-6}$&$-18^{+2}_{-2}$&$119^{+5}_{-4}$&$64^{+2}_{-2}$&$150^{+6}_{-6}$&$129^{+5}_{-4}$&$-1.39^{+0.02}_{-0.02}$&$0.37^{+0.01}_{-0.01}$&$-1.63^{+0.03}_{-0.03}$&$0.41^{+0.01}_{-0.01}$&$0.68^{+0.04}_{-0.04}$\\
	$~\,9.60-11.46$&1310&$134^{+4}_{-4}$&$-10^{+2}_{-2}$&$94^{+4}_{-4}$&$42^{+2}_{-2}$&$151^{+6}_{-6}$&$115^{+4}_{-4}$&$-1.33^{+0.02}_{-0.02}$&$0.39^{+0.01}_{-0.01}$&$-1.68^{+0.03}_{-0.03}$&$0.46^{+0.02}_{-0.02}$&$0.64^{+0.03}_{-0.03}$\\
	$11.46-13.12$&1297&$130^{+4}_{-4}$&$-6^{+2}_{-2}$&$92^{+3}_{-3}$&$38^{+2}_{-2}$&$143^{+6}_{-6}$&$109^{+4}_{-4}$&$-1.41^{+0.02}_{-0.02}$&$0.35^{+0.01}_{-0.01}$&$-1.60^{+0.03}_{-0.04}$&$0.47^{+0.02}_{-0.02}$&$0.64^{+0.03}_{-0.03}$\\
	$13.12-14.72$&1229&$119^{+4}_{-4}$&$-10^{+2}_{-2}$&$90^{+4}_{-3}$&$34^{+1}_{-1}$&$143^{+6}_{-6}$&$105^{+4}_{-4}$&$-1.39^{+0.02}_{-0.02}$&$0.32^{+0.01}_{-0.01}$&$-1.69^{+0.03}_{-0.03}$&$0.47^{+0.02}_{-0.02}$&$0.65^{+0.03}_{-0.03}$\\
	$14.72-16.46$&1201&$109^{+4}_{-4}$&$-10^{+2}_{-2}$&$84^{+3}_{-3}$&$33^{+1}_{-1}$&$140^{+6}_{-6}$&$102^{+4}_{-4}$&$-1.38^{+0.02}_{-0.02}$&$0.34^{+0.01}_{-0.01}$&$-1.67^{+0.03}_{-0.03}$&$0.41^{+0.02}_{-0.02}$&$0.68^{+0.02}_{-0.03}$\\
	$16.46-18.22$&1132&$94^{+3}_{-4}$&$-8^{+1}_{-1}$&$80^{+4}_{-3}$&$29^{+1}_{-1}$&$131^{+6}_{-6}$&$112^{+4}_{-4}$&$-1.40^{+0.01}_{-0.01}$&$0.33^{+0.01}_{-0.01}$&$-1.72^{+0.03}_{-0.03}$&$0.45^{+0.03}_{-0.02}$&$0.74^{+0.02}_{-0.02}$\\
	$18.22-20.24$&1100&$88^{+3}_{-4}$&$-7^{+1}_{-1}$&$74^{+4}_{-3}$&$27^{+1}_{-1}$&$126^{+6}_{-6}$&$98^{+4}_{-4}$&$-1.39^{+0.02}_{-0.02}$&$0.34^{+0.01}_{-0.01}$&$-1.72^{+0.03}_{-0.03}$&$0.43^{+0.02}_{-0.02}$&$0.73^{+0.02}_{-0.02}$\\
	$20.24-22.87$&1070&$76^{+4}_{-6}$&$-7^{+1}_{-1}$&$78^{+5}_{-4}$&$27^{+1}_{-1}$&$129^{+7}_{-6}$&$105^{+5}_{-4}$&$-1.42^{+0.01}_{-0.01}$&$0.30^{+0.01}_{-0.01}$&$-1.71^{+0.03}_{-0.03}$&$0.43^{+0.03}_{-0.03}$&$0.75^{+0.02}_{-0.02}$\\
	$22.87-26.79$&1047&$67^{+4}_{-6}$&$-5^{+1}_{-1}$&$70^{+6}_{-4}$&$26^{+1}_{-1}$&$141^{+6}_{-6}$&$119^{+5}_{-4}$&$-1.45^{+0.01}_{-0.01}$&$0.29^{+0.01}_{-0.01}$&$-1.68^{+0.03}_{-0.03}$&$0.46^{+0.02}_{-0.02}$&$0.71^{+0.02}_{-0.02}$\\
	$26.79-30.00$&469&$41^{+20}_{-26}$&$-4^{+2}_{-2}$&$94^{+8}_{-11}$&$27^{+2}_{-2}$&$127^{+9}_{-8}$&$99^{+6}_{-5}$&$-1.48^{+0.03}_{-0.03}$&$0.31^{+0.02}_{-0.02}$&$-1.79^{+0.05}_{-0.05}$&$0.42^{+0.04}_{-0.03}$&$0.66^{+0.04}_{-0.05}$\\
	$30.00-102.93$&984&$47^{+17}_{-29}$&$-7^{+3}_{-3}$&$85^{+11}_{-12}$&$23^{+2}_{-2}$&$124^{+6}_{-5}$&$87^{+4}_{-3}$&$-1.44^{+0.04}_{-0.03}$&$0.28^{+0.02}_{-0.02}$&$-1.79^{+0.03}_{-0.03}$&$0.38^{+0.03}_{-0.02}$&$0.48^{+0.04}_{-0.04}$\\
	\enddata
	\tablenote{Selected width of the distance in $r_\mathrm{gc}$.}
	\tablenote{$N$ is the number of stars of the observational data in the corresponding distance bin for modeling.}
\end{deluxetable*}

\subsection{Mock data test of the GMM selection method}\label{sub:test}

In this subsection, we apply a mock data test to check the accuracy of the GMM selection under different GMM parameters and $f_\mathrm{an}$. For each distance bin in Table~\ref{tab:Sgr-removed}, we use the best estimated model parameters to specify the velocity and metallicity distributions of the GMM/isotropic and the GMM/Sausage components, and then draw random samples from the specified distributions to make the mock data with one million stars by the random sampling function \texttt{Numpy.random.multivariate\_normal()} in the \texttt{Numpy} package \citep{harris2020array}. Every sampled point is convolved with a matrix of measurement errors selected randomly from the observational data. After the construction of the mock data of the GMM/isotropic and the GMM/Sausage components, we mix them with $f_\mathrm{an}$ ranging from 0.1 to 0.9. We draw the same number of fake stars as the observational data randomly from each newly constructed mock data. The GMM selection is applied to the mock data to obtain the true discovery rates of the GSE-related and the GSE-removed halo stars. We generate 200 synthetic catalogs in such manner to obtain the mean discovery rate and the standard deviation. 

Figure~\ref{fig:discover} shows the discovery rates of the GSE-related and the GSE-removed halo stars under different GMM models and $f_\mathrm{an}$. It is clear that a higher $f_\mathrm{an}$ results in a larger discovery rate of the GSE-related halo and a smaller discovery rate of the GSE-removed halo. The true value of $f_\mathrm{an}$ obtained from our K giant sample is about $0.6-0.70$, which corresponds to a typical discovery rate of $0.80-0.95$ for the GSE-related halo and $0.70-0.85$ for the GSE-removed halo. In general, it is a good result especially for the GSE-removed halo considering that the GSE is the dominant component. However, we note that the GMM is a simplified model where the metallicity distributions of the two components are simply described by a Gaussian function. Therefore, slight differences between the discovery rates of the observational and mock data may exist. 

\begin{figure}
	\centering
	\includegraphics[width=0.5\textwidth]{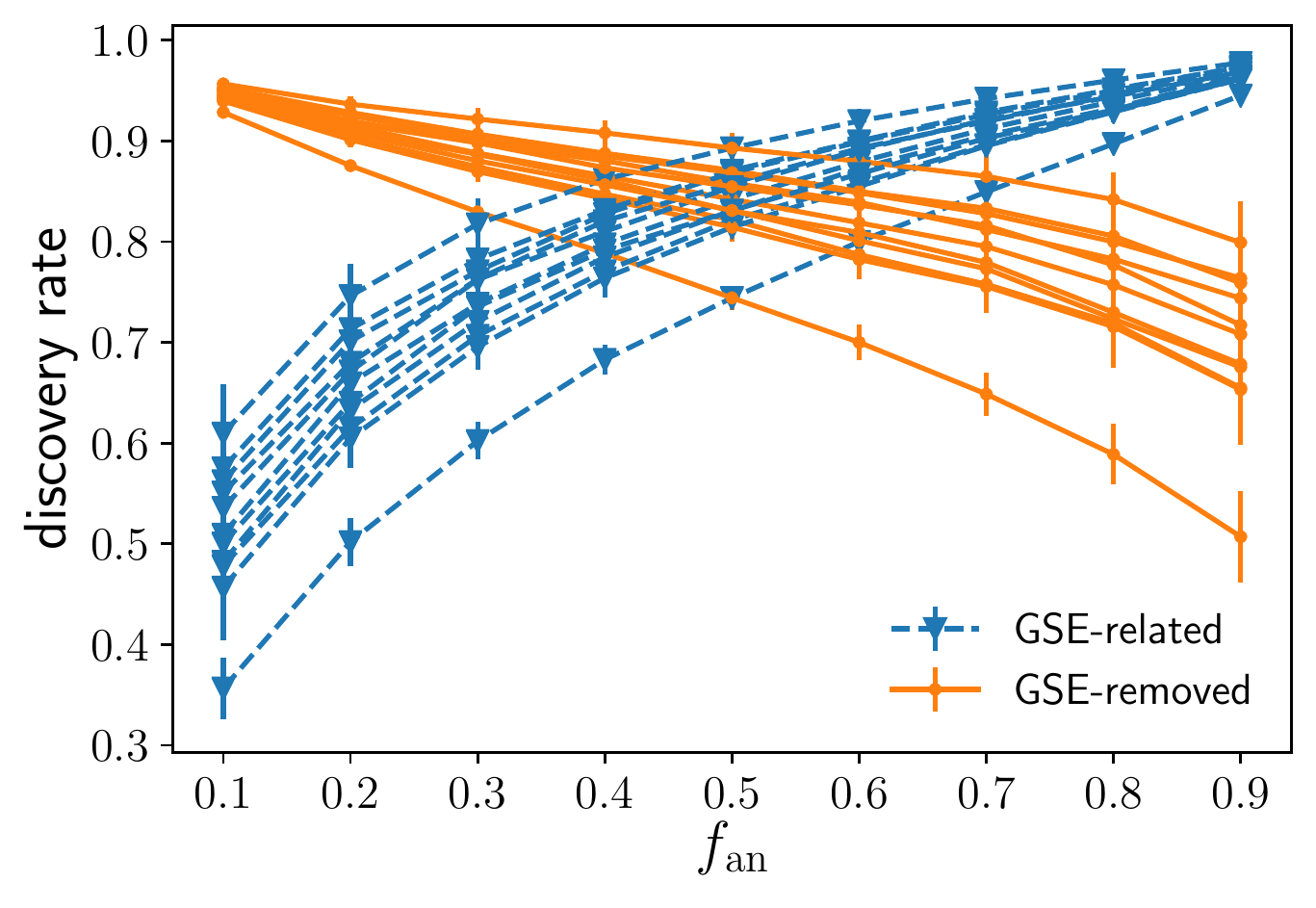}
	\caption{Discovery rates of the GSE-related (blue dash lines) and the GSE-removed (orange solid lines) stellar halos under different GMM models and $f_\mathrm{an}$ obtained by applying the GMM selection method to the mock data. Every dash and solid line represents models of the GMM/isotropic and the GMM/Sausage components of a distance bin in Table~\ref{tab:Sgr-removed}, respectively.}
	\label{fig:discover}
\end{figure}

Figure~\ref{fig:compare} shows the distributions of the spherical velocities and the metallicity of the two stellar halos selected from the observational and the mock data. These halos are consistent with the models of their corresponding GMM components, and the structure of the two $v_\mathrm{r}$ lobes of the GSE-related halo is clearly seen in distance bins with small $r_\mathrm{gc}$. We apply a Kolmogorov–Smirnov test to check the similarities between the stellar halos selected from the observational and the mock data. From the p values, we find that the GSE-related halos of the two datasets agree well with each other both in the spherical velocities and the metallicity distributions. For the GSE-removed halo, we find a good level of similarities in the spherical velocities, while some differences exist in the metallicity distribution. The GSE-removed halo has a more complex formation mechanism than the GSE-related halo, which may originate from the in-situ process, gas accretion, multiple minor mergers, and the dissolution of globular clusters, etc. Therefore, a simple Gaussian function seems to be inadequate to fit the metallicity distribution of the GMM/isotropic component well. In general, stellar halos selected from the two datasets are similar in most cases, which suggests that the discover rates obtained from the mock data can be used as a reference of the observational data. 

\begin{figure*}
	\centering
	\includegraphics[width=\textwidth]{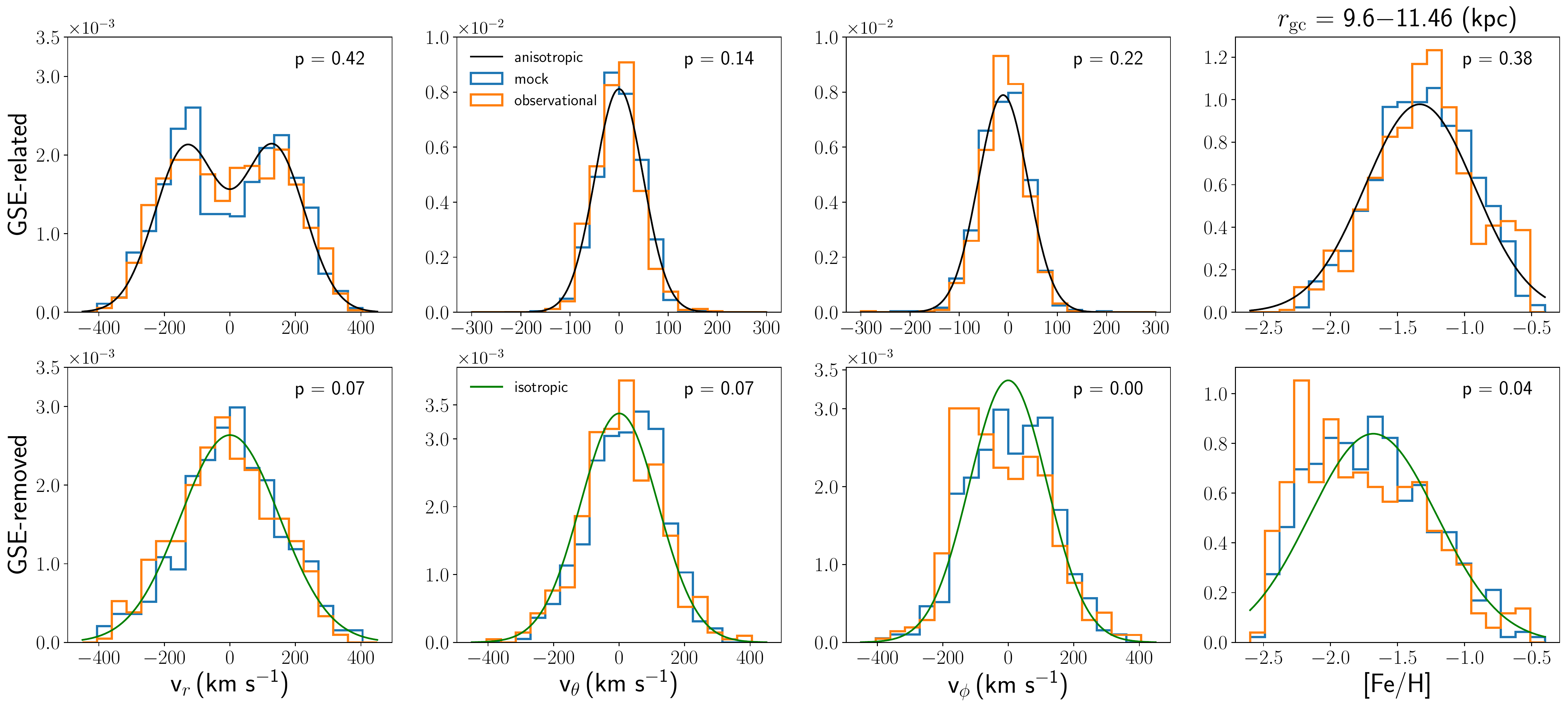}
	\includegraphics[width=\textwidth]{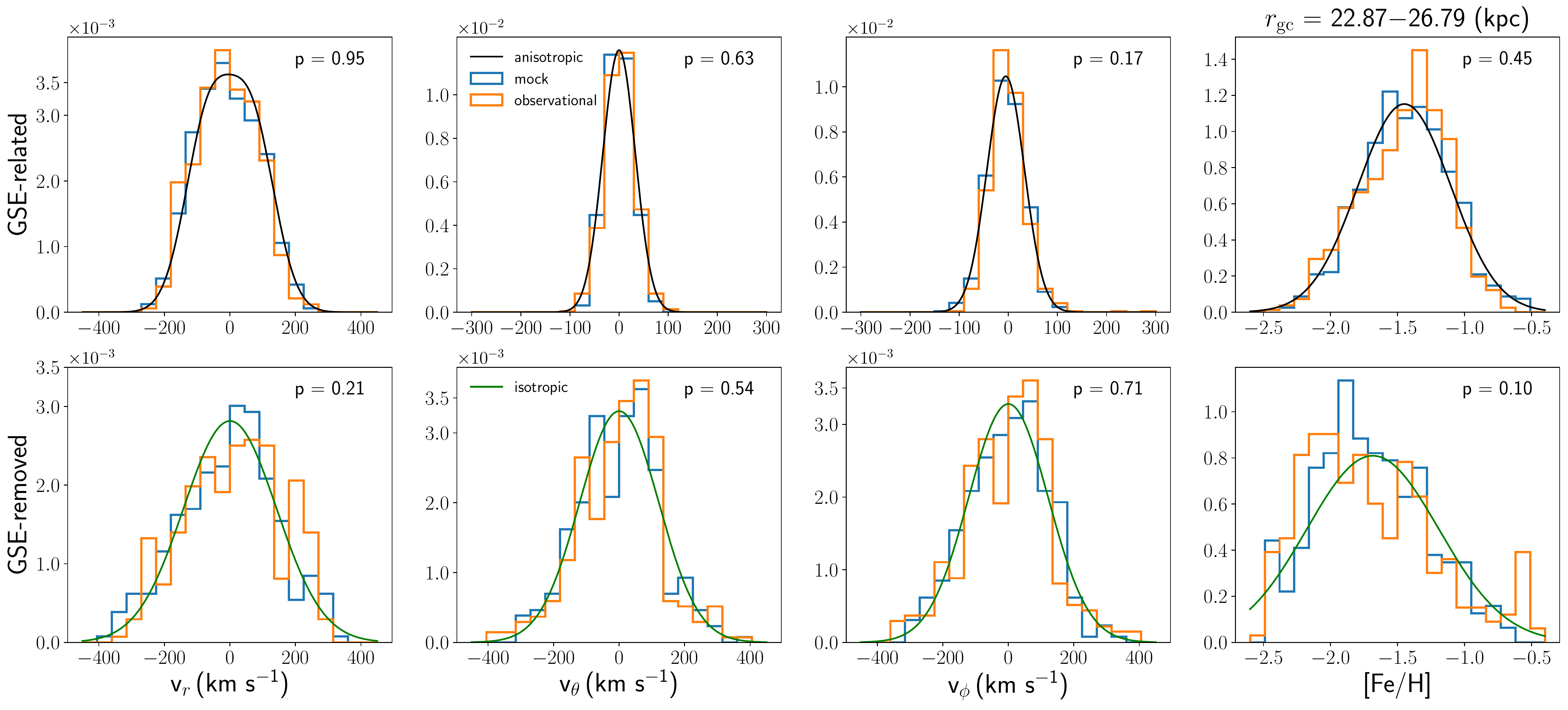}
	\caption{Comparison of the stellar halos selected from the mock data (blue histograms), the observational data (orange histograms), the GMM/anisotropic component model (black lines), and the GMM/isotropic component model (green lines) of two distance bins. The star numbers of the mock and the observational data are set to be equal. For the GSE-related halo, the numbers of the distance bins with p value larger than 0.05 are 11 for the distribution of $v_r$, 8 for the distribution of $v_\theta$, 8 for the distribution of $v_\phi$, and 10 for the distribution of [Fe/H]. For the GSE-removed halo, the numbers of the distance bins with p value larger than 0.05 are 9 for the distribution of $v_r$, 10 for the distribution of $v_\theta$, 6 for the distribution of $v_\phi$, and 4 for the distribution of [Fe/H].}
	\label{fig:compare}
\end{figure*}

\section{Fitting method}\label{sec:fit}
In this section, we describe the method used for estimating the stellar density, and apply it to derive the density shapes of the GSE-related and the GSE-removed halos. 
\subsection{Estimation of the stellar density} 

\citet{2017RAA....17...96L} and \citet{2018MNRAS.473.1244X} introduced a statistic method to derive the stellar density profile along each line of sight of a spectroscopic survey. The core of this method is to treat the contribution of a star as a probability density function extending along a line of sight. Selection effects due to the target selection and distance estimation are considered in the estimation. Using this method, we can construct a density map of the Galactic stellar halo with varying vertical flattening. We will briefly introduce the non-parametric method hereafter, and more details can be found in \citet{2017RAA....17...96L}.

We assume that the photometric survey is complete for stars brighter than the limiting magnitude.  Compared to photometric surveys, spectroscopic surveys are less complete and have a smaller coverage due to the target selecting. To correct the bias generated by selecting subsamples, we need the selection function $S$ in which the probability of a star being included in the observation is given by their Galactic coordinates ($\textit{l}, \textit{b}$), colors ($\textit{c}$), and apparent magnitudes ($\textit{m}$). For a given line of sight, the relation between the densities derived from the photometric ($\nu_\mathrm{ph}$) and the spectroscopic data ($\nu_\mathrm{sp}$) is assumed to be
\begin{equation}
    \nu_\mathrm{ph}(D|\textit{c}, \textit{m}, \textit{l}, \textit{b}) = \nu_\mathrm{sp}(D|\textit{c}, \textit{m}, \textit{l}, \textit{b}){S^{-1}(\textit{c}, \textit{m}, \textit{l}, \textit{b})},
    \label{eq:nu}
\end{equation}
where $D$ is the distance along the given line of sight.

The selection function \textit{S}(\textit{c}, \textit{m}, \textit{l}, \textit{b}) is defined as
\begin{equation}
\textit{S}(\textit{c}, \textit{m}, \textit{l}, \textit{b}) = \frac{\textit{n}_\mathrm{sp}(\textit{c}, \textit{m}, \textit{l}, \textit{b})}{\textit{n}_\mathrm{ph}(\textit{c}, \textit{m}, \textit{l}, \textit{b})},
\label{eq:selec_function}
\end{equation}
where $\textit{n}_\mathrm{sp}(\textit{c}, \textit{m}, \textit{l}, \textit{b})$ and $\textit{n}_\mathrm{ph}(\textit{c}, \textit{m}, \textit{l}, \textit{b})$ are the star counts of the spectroscopic and the photometric data in the color-magnitude diagram. The selection function of the LAMOST K giants is provided by \citet{2017RAA....17...96L}, which is based on the photometric survey of 2MASS.  

After the integration over $\textit{c}$ and $\textit{m}$, we can obtain the stellar density profile for a given line of sight as
\begin{equation}
\nu_\mathrm{ph}(D|l, b) = \iint{\nu_\mathrm{sp}(D|\textit{c}, \textit{m}, \textit{l}, \textit{b})S^{-1}(\textit{c}, \textit{m}, \textit{l}, \textit{b})\mathrm{d}c\mathrm{d}m}.
\label{eq:ph}
\end{equation}

To obtain $\nu_\mathrm{ph}$, we need to first obtain $\nu_\mathrm{sp}$ from the spectroscopic data. We apply a kernel density estimation (KDE) method to derive $\nu_\mathrm{sp}$ along a given line of sight. We treat the contribution of a star $i$ as a probability density function $p_i$ extending along the corresponding line of sight, and $p_i$ is defined as,
\begin{equation}
	p_i(D) = \frac{\mathcal{N}(D|D_i, \sigma_{D_i}^2)}{\int_{D_\mathrm{min}}^{D_\mathrm{max}}\mathcal{N}(D_x|D_i, \sigma_{D_i}^2)dD_x},
	\label{eq:pi}
\end{equation}
where $\mathcal{N}$ is a normal function. $D_i$ is the estimated distance of star i, and $\sigma_{D_i}$ is the uncertainty of the estimated distance. We set $D_\mathrm{min} = 0$ kpc and $D_\mathrm{max} = 200$ kpc.

We take into account the contribution of all stars around $\textit{c}$ and $\textit{m}$. The density profile $\nu_\mathrm{sp}$ is defined as
\begin{equation}
\nu_\mathrm{sp}(D|\textit{c}, \textit{m}, \textit{l}, \textit{b}) = \frac{1}{{\Omega}D^2}\sum_{i}^{n_{sp}(\textit{c}, \textit{m}, \textit{l}, \textit{b})}p_i(D),
\label{eq:sp}
\end{equation}
where $\Omega$ is the solid angle of the given line of sight. Since it is a constant value and has no influence on the obtained density shape, we normalize it to 1. Combining Equation~\ref{eq:ph} and ~\ref{eq:sp}, we obtain the derived $\nu_\mathrm{ph}$ as a continuous function of $D$ for a given line of sight. The LAMOST survey has thousands of plates (line of sights) to be observed, and each plate has a derived $\nu_\mathrm{ph}$. However, as \citet{2018MNRAS.473.1244X} pointed out, the derived $\nu_\mathrm{ph}$ may suffer from large uncertainties at the position with no stars Therefore, we will only adopt $\nu_\mathrm{ph}$ where we can find a star. In other words, every halo K giant is assigned a derived $\nu_\mathrm{ph}$ which represents an approximation of the stellar density at this position.

The non-parametric method requires no analytic density profiles like a Einasto function before the fitting process. The independence on any preset stellar density model enables \citet{2018MNRAS.473.1244X} to study the variation of the vertical flattening of the Galactic stellar halo. We will use the same non-parametric method as \citet{2018MNRAS.473.1244X} to obtain the density maps of the GSE-related and the GSE-removed halo samples in the following words.

\subsection{Fitting of the density shape}

We remove stars dimmer than the 2MASS limiting magnitude $K$ of 14.3 mag to ensure the completeness of the photometric survey. This final selection retains 7872 stars in the GSE-related halo and 3752 stars in the GSE-removed halo. There is no much difference in the sky area ($\textit{l}-\textit{b}$ panel) between the two stellar halos. We apply the non-parametric method to the two stellar halos and derive the density $\nu$ of their K giants, respectively. Figure~\ref{fig:map} shows the median value of $\nu$ of the K giants at each $R-|Z|$ pixel. We can see that both the GSE-related and the GSE-removed stellar halos tend to be more vertically flattened at smaller radii, and the GSE-removed halo seems to have a smaller ellipticity. 
\begin{figure*}
	\centering
	\includegraphics[width=0.8\textwidth]{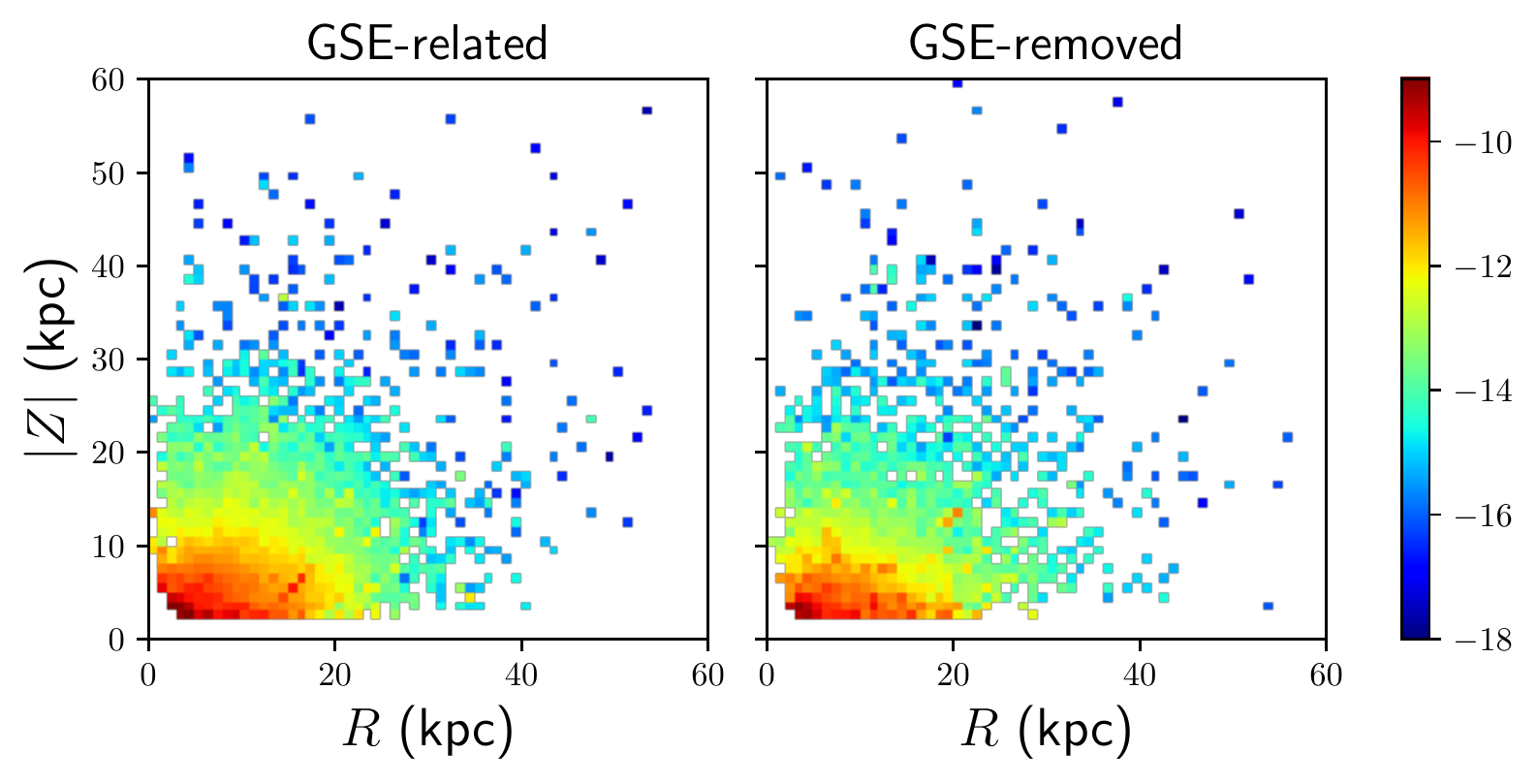}
	\caption{Number density maps of the GSE-related and the GSE-removed stellar halos. The pixel size is $1\times1$ (kpc). The number density in each pixel represents the median $\ln (\nu)$ of K giants.} 
	\label{fig:map}
\end{figure*}

In order to characterize the vertical flattening of the density shapes quantitatively, we fit the iso-density contour of the derived density $\nu$ of these K giants with a ellipsoidal profile defined as
\begin{align}
	&\nu(r) = {\nu_0}r^{-\alpha},\label{eq:spl}\\
	&r = \sqrt{R^2+(Z/q(r))^2},
\end{align} 
where $r$ is the semi-major axis of the ellipse, and $q$ is defined as the ratio of the semi-minor to the semi-major axis. In this study, we assume a ratio of the semi-intermediate to semi-major axis $p$ to be 1 ($R = X^2+Y^2/p^2, p = 1$). A recent study of the density shape of the Galactic stellar halo by RR Lyraes of \textit{Gaia} DR2 shows a $p$ value of 0.77, and the major axis is titled by $\phi$ of $21\deg$ with respect to the y-axis \citep{2019MNRAS.482.3868I}. We note that the values of $p$ and $\phi$ are obtained from the study of the whole Galactic stellar halo by RR Lyraes, which could be different for the GSE-related and the GSE-removed halos in this study. We also try to add $p$ and $\phi$ to our density model as two free parameters. However, a larger halo star sample with a more complete sky coverage is needed to obtain a reliable result of $p$ and $\phi$ as a function of $r$. Therefore, we will use the simplified density model and only focus on the variation of $q$ in this study.  

We divide the K giants into 23 ln ($\nu$) bins with a bin size of 0.25 in the range of $-15.525 < \mathrm{ln} (\nu) < -9.875$. A total of 7389 stars in the GSE-related halo and 3500 stars in the GSE-removed halo remain in this range. The specific star numbers and best estimated parameters of the two stellar halos in each ln ($\nu$) bin are recorded in Table~\ref{tab:all_results}. We fit the spatial distribution of the K giants with an ellipse in each ln $(\nu)$ bin, and then obtain the best estimated spherical radius $r$ and flattening $q$. The detailed fitting method is introduced as follows.

In the ellipsoidal coordinates, $R$ and $Z$ is defined as
\begin{align}
	& R = r\,\mathrm{cos}(\eta), \label{eq:R} \\
	& Z = r\,q\,\mathrm{sin}(\eta), \label{eq:z}\\
	&r_\mathrm{gc} = \sqrt{R^2+Z^2}. \label{eq:rgc}\\
	&\mathrm{sin}(\theta) = Z/r_\mathrm{gc} \label{eq:theta}
\end{align}

Combining Equation~\ref{eq:R}, ~\ref{eq:z}, and ~\ref{eq:rgc}, we rewrite $r_\mathrm{gc}$ as
\begin{equation}
	r_\mathrm{gc} = r\sqrt{1 + (q^2 - 1)\mathrm{sin}^2(\eta)}.
	\label{eq:rgc-q}
\end{equation}

Using Equation~\ref{eq:theta}, we write $\sin(\eta)$ as,
\begin{equation}
	\mathrm{sin}(\eta) = \frac{r_\mathrm{gc}\,\mathrm{sin}(\theta)}{rq}.
	\label{eq:eta}
\end{equation} 
 
Substituting Equation~\ref{eq:eta} into Equation~\ref{eq:rgc-q}, we can fit the iso-density contour in $r_\mathrm{gc}-\sin(\theta))$ space with Equation~\ref{eq:final},
\begin{equation}
	r_\mathrm{gc} = r q\sqrt{\frac{1}{q^2-(q^2-1)\sin^2(\theta)}}.
	\label{eq:final}
\end{equation}

\section{Results}\label{sec:result}
In this section, we show the radial density profiles of the GSE-related and the GSE-removed stellar halos. We find two Milky Way-like galaxies in TNG50 simulation, and analyze the density shapes of the major related and the major removed stellar halos.

\subsection{Number Density Shapes and Profiles of the Observational Data}\label{sub:real}

We take the fitting results of the bin with $\mathrm{ln} (\nu) = -12.25$ as an example. We fit the distributions of all K giants of $-12.50<\ln (\nu)<-12.00$ with Equation~\ref{eq:final} in $r_\mathrm{gc}- |\sin(\theta)|$ space. We use the \texttt{scipy.optimize.curve\_fit} function in the fitting. Figure~\ref{fig:fit_case} shows the best fitting results of the GSE-related and the GSE-removed halos at $\mathrm{ln} (\nu) = -12.25$. The median values of $r_\mathrm{gc}$ in each $\sin (\theta)$ bin are in accordance with the best fitting results very well. When turning to the $R-|Z|$ space, we find an ellipse of $r = \sqrt{R^2+\frac{Z^2}{q^2}}$ shown clearly by the median values and the fitting results. Both of the two stellar halos are vertically flattened at $\ln (\nu) = -12.25$, but the GSE-related halo is proven to be more spherical than the GSE-removed halo from the best estimated $q$.

The fitting results of all $\ln (\nu)$ bins are shown in Figure~\ref{fig:rgcsin_related} for the GSE-related halo and Figure~\ref{fig:rgcsin_removed} for the GSE-removed halo. In general, the distributions of the K giants in $r_\mathrm{gc}-|\sin(\theta)|$ space can be well described by Equation~\ref{eq:final}. However, due to the lack of enough data points and the large uncertainties of distance estimation in the outer halo, a certain degree of dispersion exists at severe bins with the lowest $\ln (\nu)$, for example $\ln(\nu) = -15.25$ and $\ln (\nu) = -15.62$. A larger data sample including more outer halo stars is needed to improve the accuracy of the estimation in the future. Table~\ref{tab:all_results} shows the best estimated values and the errors. We remove the outliers at $\ln (\nu) = -10.25$ in Figure~\ref{fig:sgr_relation} to ensure the reliability of the fitting.

\begin{deluxetable*}{ccccccc}
	\label{tab:all_results}
	\centering
	\tablecaption{Best estimated density shape parameters ($r$, $q$) of the K giant stellar halos.}
	\tablewidth{0pt}
	\tablehead{\multicolumn{1}{c}{}&\multicolumn{3}{|c}{GSE-related}&\multicolumn{3}{|c}{GSE-removed}}
	\decimals
	\startdata
	\hline 
	$\ln(\nu)$&Number&$r$&$q$&Number&$r$&$q$\\
	\hline
	$-10.00$&80&$11.38\pm0.53$&$0.51\pm0.04$&75&$12.56\pm1.38$&$0.38\pm0.07$\\
	$-10.25$&71&$14.26\pm0.60$&$0.40\pm0.03$&61&$16.27\pm1.54$&$0.27\pm0.04$\\
	$-10.50$&87&$13.70\pm0.46$&$0.49\pm0.03$&75&$12.66\pm0.97$&$0.45\pm0.06$\\
	$-10.75$&119&$14.55\pm0.37$&$0.49\pm0.02$&65&$13.66\pm0.57$&$0.43\pm0.03$\\
	$-11.00$&155&$15.08\pm0.39$&$0.50\pm0.02$&87&$13.05\pm0.47$&$0.54\pm0.04$\\
	$-11.25$&189&$14.81\pm0.34$&$0.60\pm0.02$&130&$14.47\pm0.43$&$0.46\pm0.03$\\
	$-11.50$&210&$15.57\pm0.31$&$0.66\pm0.02$&134&$16.24\pm0.58$&$0.44\pm0.03$\\
	$-11.75$&262&$16.47\pm0.35$&$0.67\pm0.03$&172&$16.75\pm0.54$&$0.50\pm0.03$\\
	$-12.00$&292&$18.59\pm0.40$&$0.62\pm0.02$&153&$16.63\pm0.48$&$0.56\pm0.03$\\
	$-12.25$&380&$18.26\pm0.37$&$0.68\pm0.02$&168&$18.85\pm0.54$&$0.53\pm0.03$\\
	$-12.50$&383&$19.16\pm0.38$&$0.71\pm0.02$&205&$19.97\pm0.62$&$0.53\pm0.03$\\
	$-12.75$&393&$20.22\pm0.44$&$0.74\pm0.03$&221&$19.46\pm0.69$&$0.62\pm0.03$\\
	$-13.00$&483&$20.96\pm0.46$&$0.74\pm0.03$&240&$20.39\pm0.61$&$0.64\pm0.03$\\
	$-13.25$&500&$22.27\pm0.50$&$0.76\pm0.03$&237&$23.93\pm1.16$&$0.58\pm0.04$\\
	$-13.50$&500&$23.55\pm0.52$&$0.75\pm0.03$&247&$24.73\pm1.03$&$0.61\pm0.04$\\
	$-13.75$&480&$25.73\pm0.68$&$0.75\pm0.03$&238&$24.17\pm0.76$&$0.65\pm0.03$\\
	$-14.00$&589&$26.52\pm0.75$&$0.76\pm0.03$&195&$28.50\pm1.18$&$0.65\pm0.04$\\
	$-14.25$&490&$28.30\pm0.96$&$0.78\pm0.04$&209&$28.34\pm1.31$&$0.71\pm0.05$\\
	$-14.50$&465&$27.35\pm0.83$&$0.84\pm0.04$&178&$31.71\pm2.14$&$0.65\pm0.06$\\
	$-14.75$&445&$30.65\pm1.03$&$0.79\pm0.04$&147&$32.85\pm2.05$&$0.70\pm0.06$\\
	$-15.00$&377&$31.93\pm0.83$&$0.83\pm0.06$&117&$35.81\pm1.81$&$0.74\pm0.06$\\
	$-15.25$&248&$33.07\pm2.14$&$0.94\pm0.09$&65&$41.60\pm2.94$&$0.71\pm0.07$\\
	$-15.62$&191&$39.02\pm2.54$&$0.79\pm0.08$&61&$42.27\pm2.84$&$0.79\pm0.08$
	\enddata
\end{deluxetable*}

\begin{figure*}
	\centering
	\includegraphics[width=0.8\textwidth]{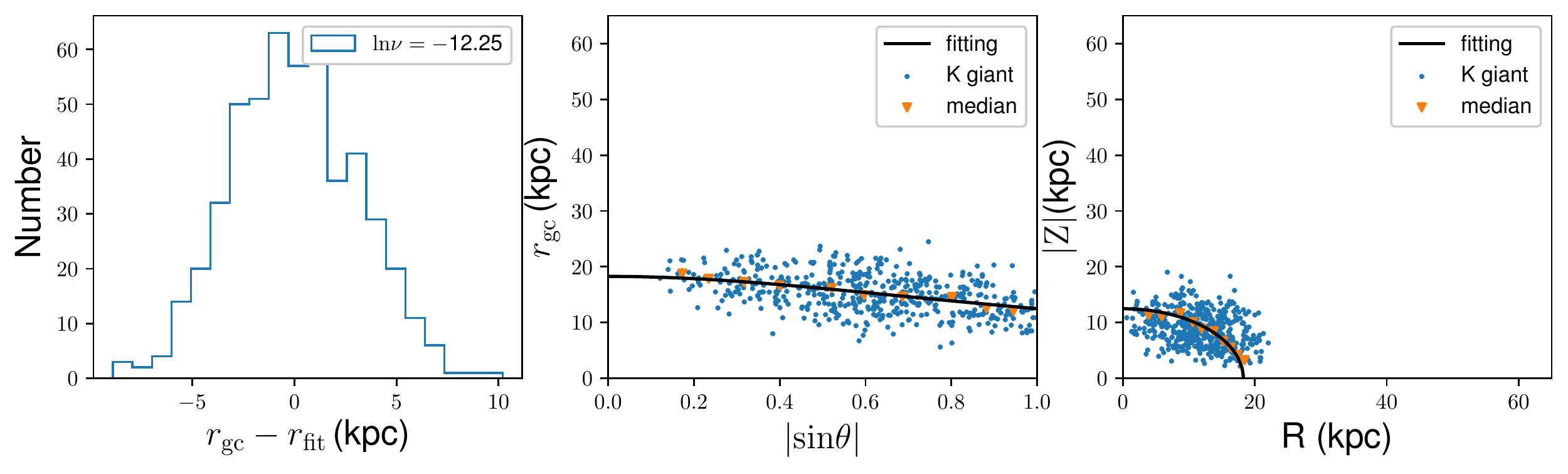}
	\includegraphics[width=0.8\textwidth]{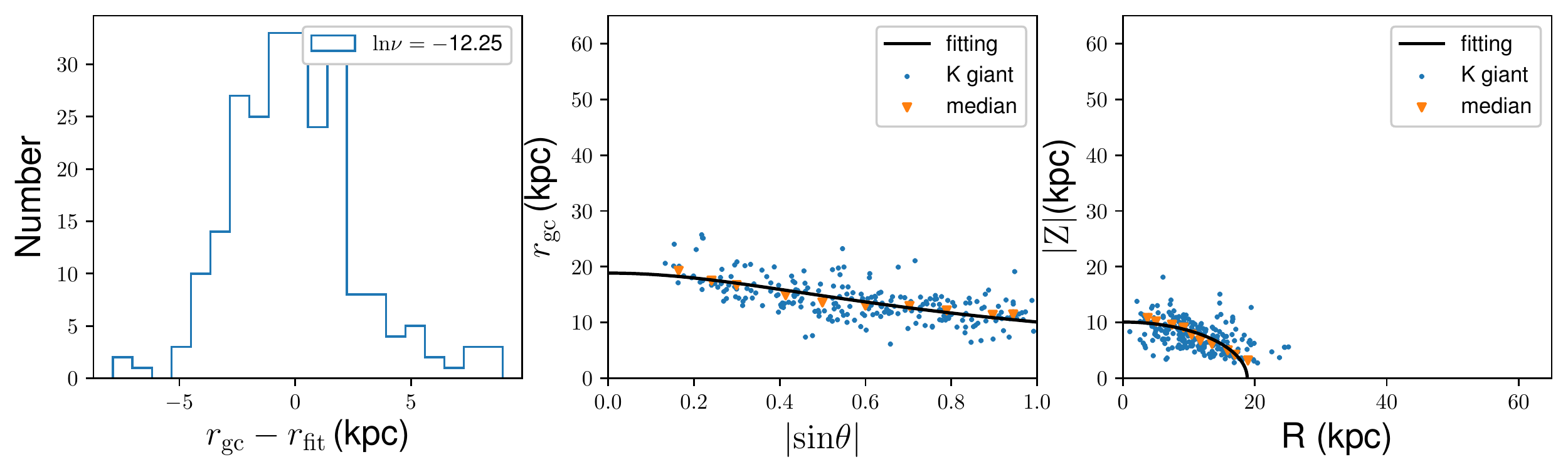}
	\caption{The fitting results of the GSE-related (upper panel) and GSE-removed (lower panel) halos at $\ln (\nu) = -12.25$. The left panel shows the histograms of the deviation of the fitting ($r_\mathrm{gc} - r_\mathrm{fit}$). The middle panel shows the distribution of the K giants (blue points) and the fitting results (black lines) in $r_\mathrm{gc}-|\sin(\theta)|$ space. The orange triangles represent the median values of $r_{gc}$ in each $\sin(\theta)$ bin, which conform to the best fitting lines very well. The right panel shows the K giants (blue points) and the fitting results (black lines) in $R-|Z|$ space. Although the distributions of the halo K giants have a large dispersion, the median values (orange triangles) and the best fitting results show a clear ellipse, which suggests that both of the two stellar halos are vertically flattened. Compared to the GSE-removed halo, the GSE-related stellar halo is less vertically at $\ln (\nu) = -12.25$.}
	\label{fig:fit_case}
\end{figure*}

\begin{figure*}
	\centering
	\includegraphics[width=\textwidth]{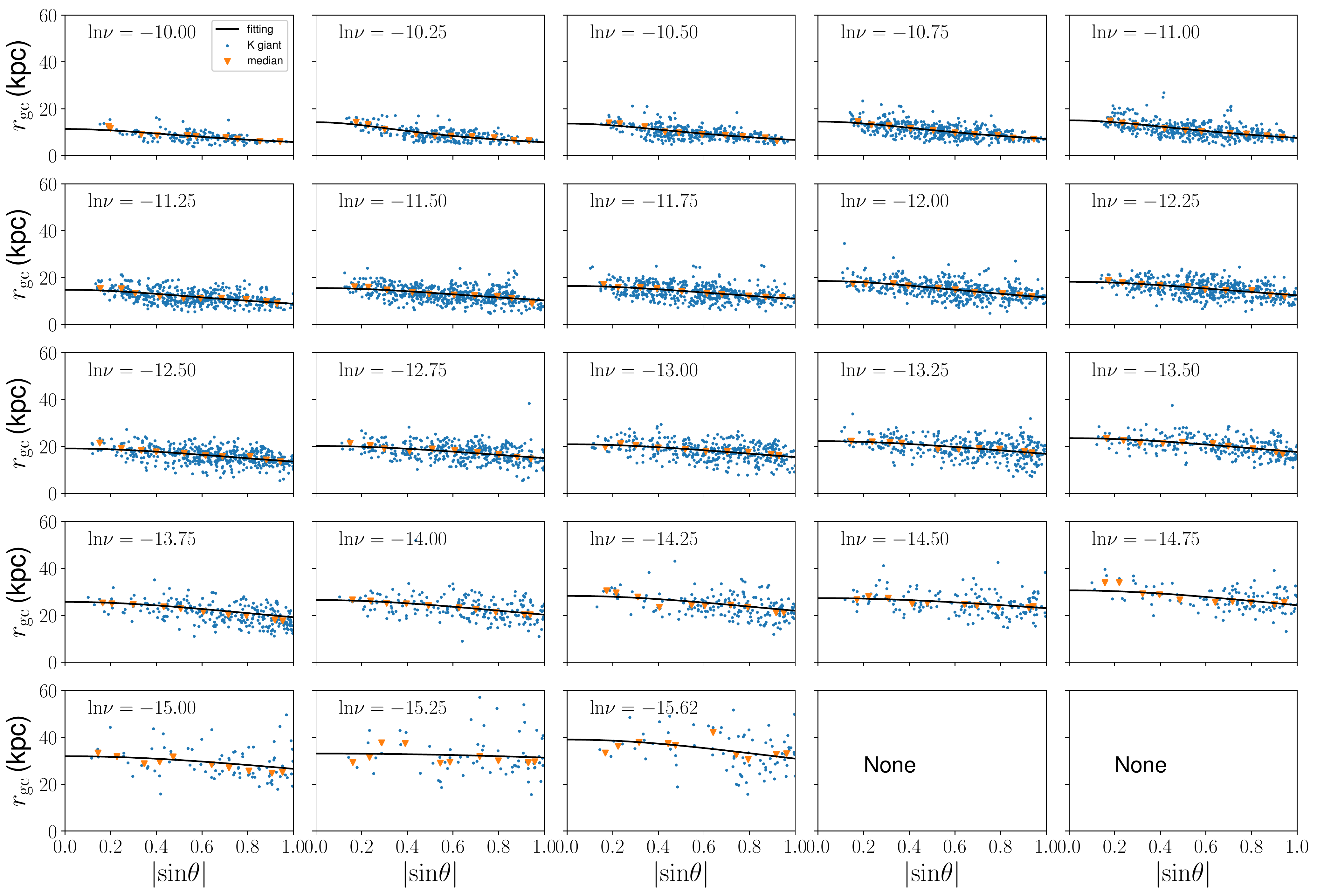}
	\caption{Fitting results of the GSE-related stellar halo in each $\ln(\nu)$ bin. The blue points represent the K giants in $r_\mathrm{gc}-|\sin(\theta)|$ space, and the black lines describe the best fitting results. The median values of $r_\mathrm{gc}$ in each $\sin(\theta)$ bin are indicated by the orange inverted triangles, and the bin size is 0.1.}
	\label{fig:rgcsin_related}
\end{figure*}

\begin{figure*}
	\centering
	\includegraphics[width=\textwidth]{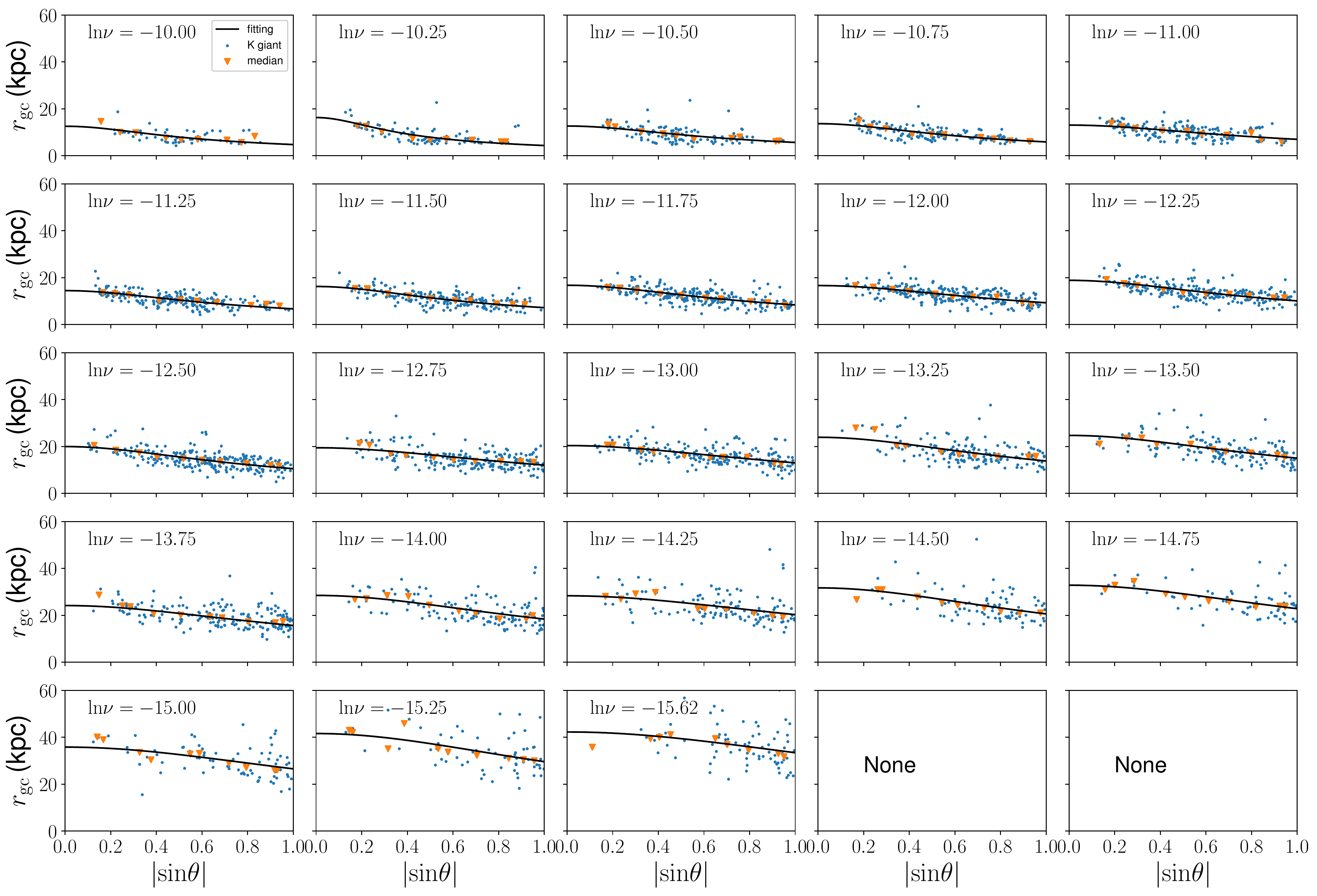}
	\caption{Fitting results of the GSE-removed stellar halo in each $\ln (\nu)$ bin. The meanings of the symbols and the lines are the same as Figure~\ref{fig:rgcsin_related}.}
	\label{fig:rgcsin_removed}
\end{figure*}

The upper panel of Figure~\ref{fig:sgr_relation} shows the variation of $q$ with the spherical radius $r$. Both of the two stellar halos are vertically flattened where $q$ ranges from 0.4 to 0.5 at the inner radii ($r < 15 \mathrm{kpc}$). The two stellar halos become more spherical with the increasing $r$. At larger radii ($r > 25 \mathrm{kpc}$), the values of $q$ are about $0.8-0.9$ for the GSE-related halo and $0.6-0.8$ for the GSE-removed halo. We define $\Delta \langle q \rangle$ ($\langle q \rangle_\mathrm{GSE-related} - \langle q \rangle_\mathrm{GSE-removed}$) as the difference of the average $q$ between the two stellar halos in a spherical radius bin with a bin size of 5 kpc. From Figure~\ref{fig:deltaq}, we find that the GSE-related halo is less vertically flattened than the GSE-removed halo, and $\Delta \langle q \rangle$ is within 0.15 in most bins. Our results is consistent with \citet{2021arXiv210810525S} that the global structure of the subsample of stars possibly associated with the GSE is more spherical. Similar cases can also be found in the stellar halos of Au8 and Au25 from Auriga simulation \citep{2019MNRAS.485.2589M}. Au8 and Au25 experienced interactions with massive satellites 4 and 0.9 Gyr ago respectively. \citet{2019MNRAS.485.2589M} found that the accreted stellar halos of Au8 and Au25 are less vertically flattened than the whole stellar halo, and the difference of $q$ is up to 0.2. The difference of $q$ between the two stellar halos indicates that the whole Galactic stellar halo may become less vertically flattened due to the GSE merger. This result corresponds to the hydrodynamical simulation that stellar halos of galaxies with a larger fraction of accreted stars from the major mergers tend to have a larger $q$ on average \citep{2021A&A...647A..95P}.

The lower panel of Figure~\ref{fig:sgr_relation} shows the relation between $\ln (r)$ and $\ln (\nu)$. We fit it with a single power law (SPL) as Equation~\ref{eq:spl}. The radial density profiles of the two stellar halos can be described well by a SPL with $\alpha = 4.97\pm0.12$ for the GSE-related halo and $\alpha = 4.25\pm0.14$ for the GSE-removed halo. 
The steeper density profile of the GSE-related halo is mainly due to the decline of the contribution of the GSE in the outer halo. We also try to fit the two density profiles with a broken power law (BPL). We obtain a $r_\mathrm{break}$ of 12.32 kpc for the GSE-related halo and 12.11 kpc for the GSE-removed halo. Considering that only one or two points are located within $r < 13$ kpc, the obtained $r_\mathrm{break}$ is not credible and more like an over-fitting. Therefore, our results are more inclined to a SPL where no $r_\mathrm{break}$ is necessarily needed. 


In this study, we divide the Galactic stellar halo into two components and find that the GSE-related halo is more spherical than the GSE-removed halo. However, a check of the non-parametric method shows that the difference of $q$ maybe exaggerated due to the different star numbers in the two stellar halos. We select 7872 stars randomly from our K giant sample (corresponding to the star number of the GSE-related halo), and 3752 stars left after the random selection (corresponding to the star number of the GSE-removed halo). We define the halo sample with 7872 stars as the halo-7872, and the other one is referred to as the halo-3752. We then apply the non-parametric method to the two halo samples and obtain $\Delta \langle q \rangle$ ($\langle q \rangle_\mathrm{halo-7872} - \langle q \rangle_\mathrm{halo-3752}$) in each $r$ bin. We repeat such manner 200 times and find that the halo-7872 tends to be less vertically flattened than the halo-3752 on average. This bias is mainly caused by the special treatment used in Equations~\ref{eq:ph}, \ref{eq:pi}, and~\ref{eq:sp}. In Equation~\ref{eq:ph}, every K giant is assigned a derived $\nu_\mathrm{ph}$ which includes the contribution of this K giant and other stars in the same line of sight. The decrease of the star number reduces the contribution of other stars and causes the decline of $\nu_\mathrm{sp}$, which will further lead to the decline of $\nu_\mathrm{ph}$. The calculated density shape will remain unchanged if the decline of $\nu_\mathrm{ph}$ is the same everywhere. However, due to the vertical flattening, stars in our halo sample tend to be closer to each other in a line of sight pointing to a higher galactic latitude, which makes the contribution of other stars more important. Therefore, the decrease of star number tends to cause a larger decline of $\nu_\mathrm{sp}$ and $\nu_\mathrm{ph}$ in a line of sight pointing to a higher latitude. We define the median $\Delta \langle q \rangle$ of the halo-7872 and the halo-3752 as the signal introduced by the different star numbers, and the standard deviation as the 1 $\sigma$ error. The median $\Delta \langle q \rangle$ caused by the different star numbers is about $0.03-0.06$ as shown in Figure~\ref{fig:deltaq}, which may partly contribute to the $\Delta \langle q \rangle$ of the GSE-related and the GSE-removed halos. To reduce this bias as possible, we select a subsample of 3752 stars randomly from the GSE-related halo and compare it with the GSE-removed halo. We repeat such manner 200 times and obtain the median $\Delta \langle q \rangle$ ($\langle q \rangle_\mathrm{subsample} - \langle q \rangle_\mathrm{GSE-removed}$) in each $r$ bin. In Figure~\ref{fig:deltaq}, we find that this median $\Delta \langle q \rangle$ slowly increases from 0.04 to 0.07 at $r \sim 10-30$ kpc and suddenly increases up to 0.14 at $r \sim 30-35$ kpc. Although it still supports the view that the GSE-related halo is less vertically flattened than the GSE-removed halo, the intrinsic difference of $q$ between the two stellar halos is probably less obvious than the result shown in Figure~\ref{fig:sgr_relation}. The long-time dynamical evolution of the GSE members in the Galactic halo may erase their features in the spatial distribution, which makes the GSE-related halo less distinguishable from $q$ than the accreted stellar halos of Au8 and Au25. 
\begin{figure}
	\centering
	\includegraphics[width=0.5\textwidth]{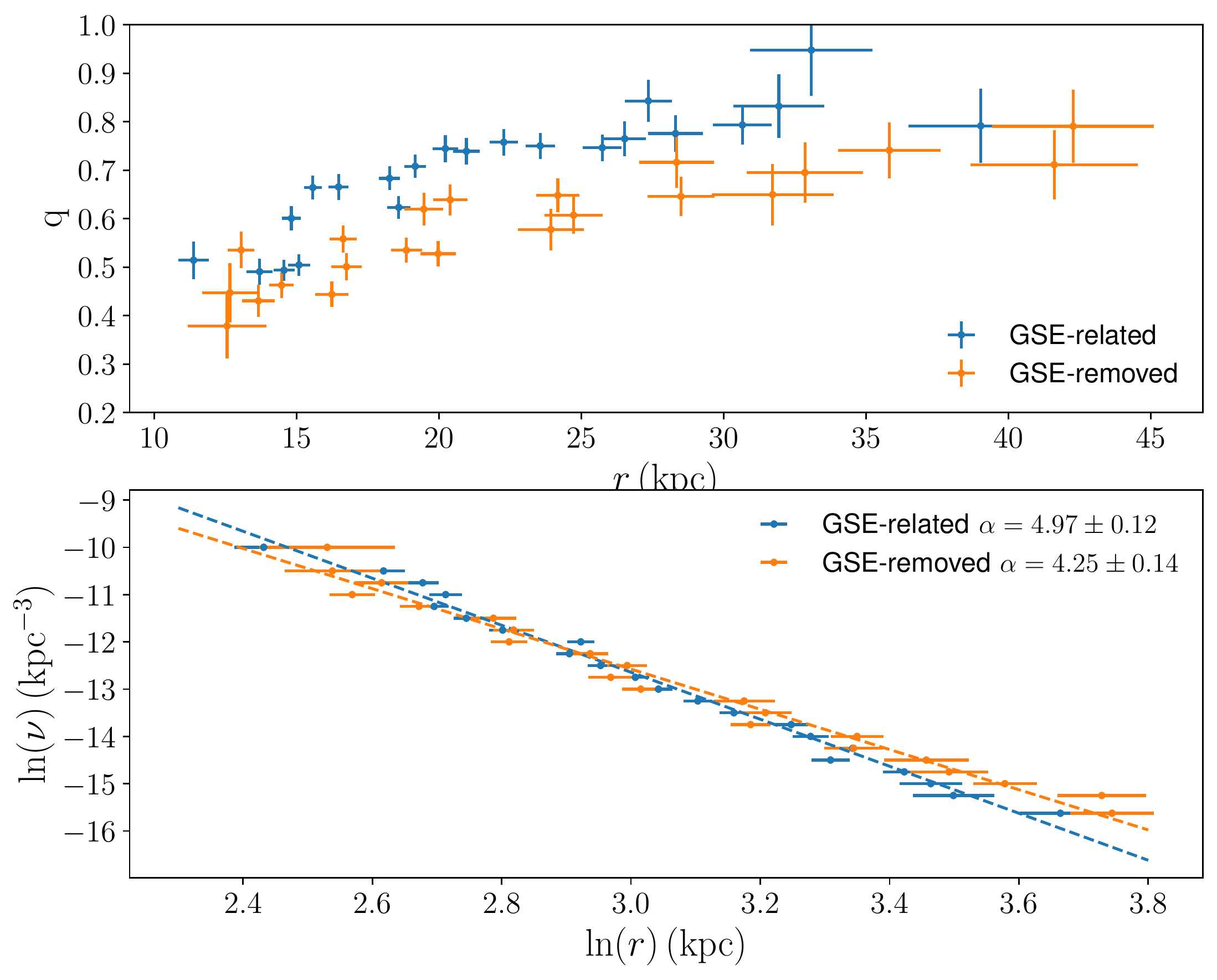}
	\caption{Upper panel shows the relations between $r$ and $q$ for the GSE-related (blue cross) and the GSE-removed (orange cross) halos. We find that both the two halos are highly vertically flattened at the inner radii but become more spherical at the outer radii. The GSE-related halo is more spherical than the GSE-removed halo. The relations between $\ln (r)$ and $\ln (\nu)$ are shown in the lower panel. We find that the radial density profiles of the two stellar halos can be fitted by a single power law (dash lines) of $\nu = \nu_0r^{-\alpha}$ well. The index $\alpha$ is $4.97\pm0.12$ for the GSE-related halo and $4.25\pm0.14$ for the GSE-removed halo.}
	\label{fig:sgr_relation}
\end{figure}


\begin{figure}
	\centering
	\includegraphics[width=0.5\textwidth]{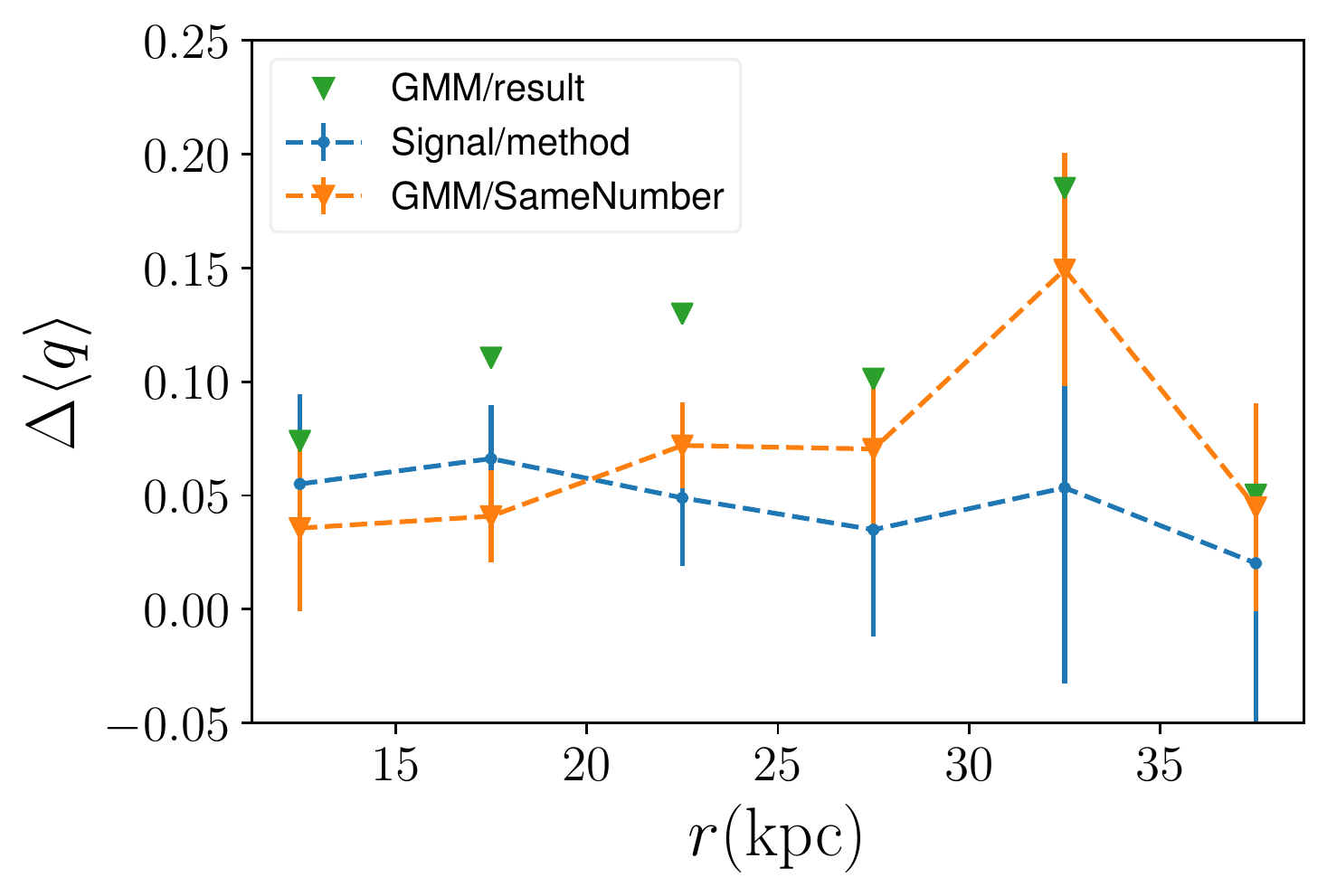}
	\caption{Mean differences $\Delta \langle q \rangle$ in each spherical radius bin for the GMM result ($\langle q \rangle_\mathrm{GSE-related} - \langle q \rangle_\mathrm{GSE-removed}$, green triangles), the signal introduced by the different star numbers ($\langle q \rangle_\mathrm{halo-7872} - \langle q \rangle_\mathrm{halo-3752}$, blue crosses), and the subsample result of the GMM ($\langle q \rangle_\mathrm{subsample} - \langle q \rangle_\mathrm{GSE-removed}$, orange crosses). The GSE-related halo is more spherical than the GSE-removed halo, but the difference of $q$ shown in Figure~\ref{fig:sgr_relation} may be exaggerated due to the different star numbers of the two stellar halos.}
	\label{fig:deltaq}
\end{figure}
\subsection{Milky Way Analogues in TNG50}\label{sub:simulation}

The lack of enough observational data prevents us from obtaining a full morphology of the accreted stellar halo. In this subsection, we select two Milky Way analogues that experience a significant major merger event from the TNG50 simulation, and study the density shapes of the major merger-related and the major merger-removed stellar halos. Using cosmological hydrodynamical simulations can help us disentangle the impact of the major merger event on the stellar halo without the bias caused by the selection effect and distance estimation. 

The TNG50 is the highest resolution volume in the IllustrisTNG cosmological hydrodynamical simulations \citep{2018MNRAS.473.4077P,2019ComAC...6....2N}. The simulation follows the evolution of $2\times2160^3$ initial resolution elements. The mass of the dark matter (DM) particle is $4.5\times10^5 M_\odot$, and the mean mass of the gas and the star particles is $8.5\times10^4 M_\odot$. 


\citet{2021ApJ...913...36E,2021ApJ...918....7E} selected a sample of 25 Milky Way like galaxies by the total subhalo mass and the mass fraction of stars satisfying $\epsilon > 0.7$ from the TNG50 simulation. All of the selected galaxies are required to have a disklike, rotationally-supported morphology as the Milky Way. We first check the merger histories of those 25 selected Milky Way like galaxies by their merger trees \citep{2015MNRAS.449...49R}. We require that the host galaxy had experienced at least one major merger with a satellite having a stellar mass larger than $10^9 M_\odot$ at redshift $z < 4$. We remove the galaxy with ID 529365 because its major merger related stars are mainly located within $r_\mathrm{gc} < 5$ kpc. We finally obtain two Milky Way analogues with IDs 506720 and 505586 at redshift $z = 0$. Galaxy 505586 experienced a merger with galaxy 231738 (ID at the snapshot of 42 in the TNG50 simulation) of a stellar mass of $2.4 \times 10^{10}\,M_\odot$ 9.0 Gyr ago, and galaxy 506720 experienced a merger with galaxy 270761 (ID at the snapshot of 46) of a stellar mass of $1.0 \times 10^{10}\,M_\odot$ 8.4 Gyr ago. We define stars originating from the massive satellite of the last major merger as the major related stellar component, and the other stars as the major removed stellar component. 

We use a kinematic definition of the stellar halo as in previous numerical works \citep{2013MNRAS.434.3142A,2017MNRAS.465.3446G,2019MNRAS.485.2589M}. We first define the net rotation direction $\mathbf{j}_\mathrm{net}$ (z-axis) as
\begin{equation}
    \mathbf{j_\mathrm{net}} \equiv \frac{\mathbf{J}_\mathrm{tot}}{M_\mathrm{tot}} = \frac{\sum_{i}m_{i}\mathbf{r}_{i} \times \mathbf{v}_i}{\sum_i m_i},
\end{equation}
Where $i$ refers to a star particle in the host galaxy. The orbital circularity parameter $\epsilon_i$ for a star $i$ is defined as
\begin{equation}
    \epsilon_i = \frac{J_{\mathrm{z},i}}{J_c(E_i)},\,\,J_c(E_i) = \sqrt{GM(<r_c)r_c},
\end{equation}
where $J_{\mathrm{z},i}$ is the magnitude of angular momentum along the z-axis, and $J_c(E_i)$ is the maximum specific angular momentum at the same specific binding energy $E_i$. A more detail introduction of the circular radius $r_c$ and binding energy $E_i$ can be found in \citet{2021ApJ...913...36E}. We keep the subset of star particles satisfying $\epsilon_i < 0.7$ as the spheroidal component, and view it as the stellar halo.  

The completeness of the simulation data allows us to study the full morphology of the stellar halo including the semi-intermediate to semi-major axis ratio $p$. Motivated by \citet{2021ApJ...913...36E}, we calculate the stellar density shape using a local iterative shell method (LISM). We split the stellar halo into 100 logarithmic radial thin shells with the spherical radius $r_\mathrm{sph}$ ranging from $r_\mathrm{sph}^\mathrm{min} = 5$ to $r_\mathrm{sph}^\mathrm{max} = 40$ or 60 kpc. The $r_\mathrm{sph}^\mathrm{min} = 5$ kpc is used to eliminate the bulge stars. We assume that the spatial distribution of star particles in each shell follows a triaxial ellipsoid. For a given $r_\mathrm{sph}$, the axis lengths and the orientations are the eigenvalues and the eigenvectors of the reduced inertia tensor defined as
\begin{equation}
I_{ij} \equiv (\frac{1}{M_\mathrm{tot}}) \times \sum_{n=1}^{N} \frac{m_{n}x_{n,i}x_{n,j}}{R_n^2 (r_\mathrm{sph})},\,\,i,j = 1, 2, 3,
\end{equation}
where $M_\mathrm{tot} = \sum_{n=1}^{N} m_n$ and $N$ is the total number of star particles in this thin shell. Symbol $m_n$ represents the mass of the $n$th particle, and $x_{n,i}$ is the $i$th coordinate of the $n$th particle. The reduced inertia tensor has been widely used in calculating the density shapes of DM and stellar components in previous numerical works \citep{2002ApJ...574..538J,2012JCAP...05..030S,2021MNRAS.504.6033S}. The normalization of the particle positions, $R_n(r_\mathrm{sph})$, is the elliptical of the $n$th particle defined in terms of the halo axis length
\begin{equation}
	R_n^2(r_\mathrm{sph}) = \frac{x_n^2}{a^2(r_\mathrm{sph})} + \frac{y_n^2}{b^2(r_\mathrm{sph})} + \frac{z_n^2}{c^2(r_\mathrm{sph})},
\end{equation}
where $(a, b, c)$ represent the axis length of the triaxial halo. We require that star particles in this thin shell satisfy $0.75 \leq R_n^2 \leq 1$. At every $r_\mathrm{sph}$, we start with $a = b = c = r_\mathrm{sph}$ in the first iteration. Then we use the eigenvalues and eigenvectors of the reduced inertia tensor to deform the above shell while keeping the volume within the ellipsoid fixed. The new axis lengths $(a, b, c)$ are rescaled and given by
\begin{align*} 
a &= \frac{r_\mathrm{sph}}{(abc)^{\frac{1}{3}}} \sqrt{\lambda_1}, \\
b &= \frac{r_\mathrm{sph}}{(abc)^{\frac{1}{3}}} \sqrt{\lambda_2}, \\
c &= \frac{r_\mathrm{sph}}{(abc)^{\frac{1}{3}}} \sqrt{\lambda_3}, \label{eq:lambda}
\end{align*}
where $\lambda_i (i = 1, 2, 3; \lambda_1 \geq \lambda_2 \geq \lambda_3)$ are the absolute eigenvalues of the reduced inertia tensor. In the $m$th iteration, we define the intermediate to major axis ratio $p_m$ as $\frac{b_m}{a_m}$ and minor to major axis ratio $q_m$ as $\frac{c_m}{a_m}$. The iteration algorithm is terminated when either the residual of the shape parameters converge to a given tolerance (Max $[((p_m - p_{m-1})/p_m)^2, ((q_m - q_{m-1})/q_m)^2] \leq 10^{-3}$) or the total number of star particles inside the shell is smaller than 1000. 

An advantage of the simulation data is that we can directly select stars from the massive satellite by the provided particle IDs. We apply the LSIM to the major related and the major removed stellar halos of the two Milky way analogues. The density shape parameters $(p, q)$ at redshifts z = 0 are shown as functions of $r_\mathrm{sph}$ in Figure~\ref{fig:analogues}. The main difference is the value of $q$. It is evident that the major related stellar halo is less vertically flattened than the major removed stellar halo. For the galaxy 505586, the difference of $q$ peaks at $r_\mathrm{sph} = 5$ kpc with a value of 0.35. Then it declines progressively with the increasing $r_\mathrm{sph}$, and finally keeps in the range of 0.11 to 0.18 at $r_\mathrm{gc} > 17$ kpc. This tendency is mainly caused by the decrease of $q$ in the major related stellar halo, while $q$ of the major removed halo hardly changes. Things are different for the galaxy 506720, where the major removed stellar halo has a larger variety of $q$ than the major related stellar halo. The difference of $q$ is within 0.1 at $r_\mathrm{gc} = 10 - 29$ kpc, which is much smaller than the galaxy 505586 and more similar to the Milky Way. At the outer halo, this difference increases with a peak around 0.18 at $r_\mathrm{gc} ~ 33$ kpc, and finally keeps in the range of 0.10 to 0.15 at $r_\mathrm{gc} = 40 - 60$ kpc. Our study of the two Milky Way analogues also suggests that the global structure of stars associated with the major merger event is less vertically flattened than the other stars.


\begin{figure*}
	\centering
	\includegraphics[width=0.8\textwidth]{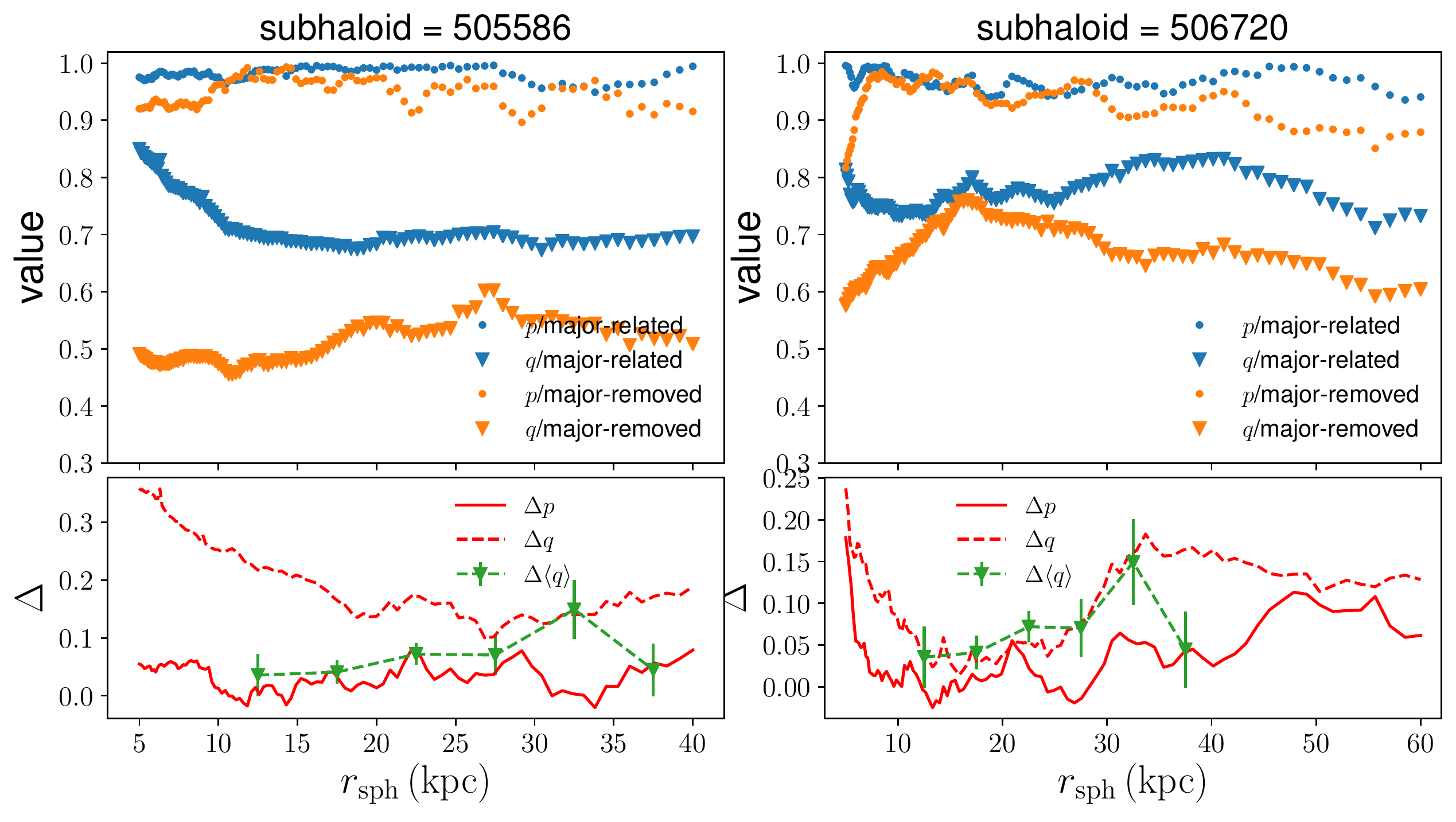}
	\caption{Upper panel shows the density shape parameters $(p, q)$ as functions of $r_\mathrm{sph}$ for the major related (blue points and triangles) and the major removed (orange points and triangles) halos of the two Milky Way analogues at redshifts z = 0. Bottom panel shows the difference $(\Delta p, \Delta q)$ of the density shape parameters between the two stellar halos. Mean difference $\Delta 
		\langle q \rangle$ (green crosses) for the subsample result of the GMM ($\langle q \rangle_\mathrm{subsample} - \langle q \rangle_\mathrm{GSE-removed}$) in Figure~\ref{fig:deltaq} is added as a comparison. We can see that the Milky Way and galaxy 506720 share a similar difference in vertical flattening at $r_\mathrm{sph} = 10 - 35$ kpc.}
	\label{fig:analogues}
\end{figure*}


\section{Summaries}\label{sec:summary}
In this study, we dived the halo K giants into two components of the GSE-related and the GSE-removed stellar halos by the GMM. The radial density profiles of the two stellar halos are determined through a non-parametric method which allows the variation of the vertical flattening with the spherical radius. Our results are compared with two Milky Way analogues which are significantly influenced by the last major merger. We summarize our main results as follows.

1. Both of the two stellar halos are vertically flattened at the inner radii and become more spherical with the increasing spherical radius. The GSE-related stellar halo is less vertically flattened than the GSE-removed halo, which suggests that the Galactic stellar halo may become more spherical due to the last major merger. The difference of $q$ between the two stellar halos is about $0.07-0.15$. After considering the possible bias caused by the different star numbers, this difference is thought to be within 0.08 in most of the spherical radii expect for $r = 30 - 35$ kpc.  

2. The radial density profiles of the two stellar halos can be fitted well with a SPL assuming that the halo density shape varies with the spherical radius. The index $\alpha$ of the SPL is $4.92\pm0.12$ for the GSE-related halo and $4.25\pm0.14$ for the GSE-removed halo. The steeper density profile of the GSE-related stellar halo is mainly caused by the decline of the fraction $f_\mathrm{an}$ in the outer halo.

3. Using LSIM method, we study the stellar density shape parameters ($p, q$) of the two Milky Way analogues in the TNG50 simulation. Our result shows that the major related stellar halo is less vertically flattened than the major removed stellar halo for the two analogues. 

In this study, we explore the influence of the GSE on the stellar halo by comparing the density shapes of the GSE-related halo with the GSE-removed halo. The difference in vertical flattening between the two stellar halos is exaggerated due to the signal introduced by the special treatment in the non-parametric method. We suspect that this bias will be reduced by making use of a halo star sample with a larger star number and more complete sky coverage. The lack of enough data points also prevents us from determining the semi-intermediate to semi-major axis ratio and orientation together. Future work will disentangle the influence of the GSE on the global structure of the Galactic stellar halo, using better data from the further data release of the Gaia and other ground based spectroscopic surveys like the SDSS-\uppercase\expandafter{\romannumeral5} and 4MOST Consortium Survey \citep{2017arXiv171103234K,2019Msngr.175...23H}.

\acknowledgements
 This study is supported by the National Natural Science Foundation of China under grant Nos. 11988101, 11890694, and 11873052 and National Key R$\&$D Program of China No. 2019YFA0405500. Guoshoujing Telescope (the Large Sky Area Multi-Object Fiber Spectroscopic Telescope LAMOST) is a National Major Scientific Project built by the Chinese Academy of Sciences. Funding for the project has been provided by the National Development and Reform Commission. LAMOST is operated and managed by the National Astronomical Observatories, Chinese Academy of Sciences. This work presents results from the European Space Agency (ESA) space mission Gaia. Gaia data are being processed by the Gaia Data Processing and Analysis Consortium (DPAC). Funding for the DPAC is provided by national institutions, in particular the institutions participating in the Gaia MultiLateral Agreement (MLA). This publication makes use of data products from the Two Micron All Sky Survey, which is a joint project of the University of Massachusetts and the Infrared Processing and Analysis Center/California Institute of Technology, funded by the National Aeronautics and Space Administration and the National Science Foundation.


\bibliography{sample63}
\bibliographystyle{aasjournal}



\end{document}